\documentstyle[12pt]{article}
\def\beq{\begin{equation}}
\def\eeq{\end{equation}}
\def\bea{\begin{eqnarray}}
\def\eea{\end{eqnarray}}
\def\bq{\begin{quote}}
\def\eq{\end{quote}}
 
\begin{document}
\pagestyle{empty}
\begin{flushright}
\end{flushright}
\vspace*{5mm}
\begin{center}
{\bf SOLVING LOOP EQUATIONS BY HITCHIN SYSTEMS VIA HOLOGRAPHY IN LARGE-$N$
 $QCD_4$ }
\\  
\vspace*{1cm} 
{\bf M. Bochicchio} \\
\vspace*{0.5cm}
INFN Sezione di Roma \\
Dipartimento di Fisica, Universita' di Roma `La Sapienza' \\
Piazzale Aldo Moro 2 , 00185 Roma  \\ 
e-mail: marco.bochicchio@roma1.infn.it \\
\vspace*{2cm}  
{\bf ABSTRACT  } \\
\end{center}
\vspace*{5mm}
\noindent
 For (planar) closed self-avoiding loops we construct a "holographic" map from
 the loop equations of large-$N$ $QCD_4$ to an effective action defined over
 infinite rank Hitchin bundles. We compute the effective action in a closed form
 in terms of functional determinants. The construction works as follows.  
 Large-$N$ $QCD_4$ on $T^2 \times T^2$ $(T^2 \times T_{\theta}^2)$ is equivalent
 by Eguchi-Kawai to a partially quenched (twisted) theory on $T^2$ described
 non-canonically by Hitchin pairs $(A=A_z dz+ A_{\bar z} d{\bar z},\Psi=-iD_u dz
 +i \bar{D_u} d{\bar z})$ with $A$ transforming as a connection and $\Psi$ as a
 Higgs field with respect to the gauge group on $T^2$. (Centrally extended)
 parabolic Hitchin bundles classify a dense set of orbits of pairs $(A,\Psi=\psi+ \bar \psi)$
 obtained as hyper-Kahler quotient by the action of the gauge group.
 We study the loop equations of the quenched (twisted) theory for the
 non-Hermitian connection $B=A+i(\kappa \psi+\bar \kappa  \bar \psi);\kappa \ne 0$ 
 and planar loops.
 We change variables in the loop equations from (centrally extended) parabolic
 Hitchin bundles to the corresponding (centrally extended) holomorphic 
 (i.e. chiral) de Rham local systems.
 The key point is that the quantum contribution to the loop equations for the
 holomorphic de Rham local systems can be evaluated by residues in a certain
 regularization and turns out to depend on the loop orientation and self-intersection.
 The outcome is that, for self-avoiding loops, the original loop equations are implied by
 the critical equation of an effective action (the master equation)
 defined over the microcanonical ensemble of the inductive limit of finite
 dimensional gauge orbits in the quenched theory, or over infinite dimensional
 orbits, arising from a central extension of Hitchin bundles, in the twisted 
 one. Finally, on the basis of the geometric Langlands
 correspondence, we propose a dual construction in which the moduli of Hitchin
 bundles are quantised and become differential operators. The loop equations are
 then implied by the condition that the master equation annihilates the vacuum
 and the master field is realized as an element of a quasi-conformal (vertex)
 algebra.
\vspace*{1cm}
\begin{flushleft}
\end{flushleft}
\phantom{ }
\vfill
\eject

\setcounter{page}{1}
\pagestyle{plain}

\section{Introduction.}

To say it in a nutshell, in this paper we construct an effective action whose
critical points solve the loop equations \cite{MM,MM1} of large-$N$ $QCD$ in four 
dimensions \cite{H4} for (planar) closed self-avoiding loops.\\ 
Being a functional over Hitchin systems \cite{Hi,Hi1}(see appendix B), our effective
action is, in a sense, the analogue for large-$N$ $QCD_4$ of the Seiberg-Witten low energy
effective action for the Abelian branch of $ {\cal N}=2$ $SUSY$ gauge theories in
four dimensions \cite {SW}. However, we should stress that our effective 
action is derived directly by the loop equations of the large-$N$ pure gauge
theory in four dimensions for (planar) closed self-avoiding loops, once carefully
chosen variables are employed in the functional integral.\\
The effective action is derived via a "holographic" map \cite{Mal,Wh,Gu}, that 
allows us 
to map the loop equations, that live on the boundary, i.e. on the loops,
into a master equation, that is the critical equation of an effective action
defined on the bulk, i.e. on the space-time.\\
One key feature responsible for the existence of the "holographic" map is a
"chiral" change of variables, that allows us to 
reduce the computation of the quantum contribution in the loop equations to the
evaluation of a residue, that depends only on the loop orientation for
self-avoiding loops. \\
Another important feature of our "holographic" map is that the residue,
that arises from the contour integral along planar self-avoiding loops
in the loop equations, is cancelled by a $u(1)$ anomalous
contribution from the "chiral" change of variables in the bulk, provided the
conformal factor of the metric is suitably chosen. \\
Despite the holographic suggestions, in this introduction we would like to make
contact with the concept of master field \cite{W}, that would occur as the solution of
the master equation for our effective action, with special regard to some
algebraic features that arose in the subject. \\
From a purely algebraic point of view, it is of the utmost importance that the
algebra generated by the monodromy of the connection that solves the loop
equations for closed self-avoiding loops be hyper-finite, i.e. be the weak limit
of a sequence of finite dimensional von Neumann algebras (see appendix A),
a property shared (trivially) by the algebra generated by the distribution of
the eigenvalues that solves the large-$N$ limit of some matrix models \cite{PAR}.
For the existence of an effective action defined over a hyper-finite
algebra (a HF effective action in short) a crucial role is played 
by the condition that closed loops be self-avoiding in the loop equations.
Indeed it has been known for some time that in the general case of
arbitrary self-intersection, the ambient algebra, in which the connection
that solves the loop equations lives, is (algebraically isomorphic to)
a free group factor with an infinite number of generators (see appendix A),
that is not a hyper-finite von Neumann algebra.
The lack of hyper-finiteness seems to be one of the reasons that so far 
prevented an exact treatment of the general case. \\ 
The existence in large-$N$ $QCD_4$ of an hyper-finite effective action
for (planar) closed self-avoiding loops, as opposed to the general case, has a
precise analogue in large-$N$ $QCD_2$ that is explicitly solvable \cite{KK}.
In the two dimensional case the von Neumann algebra generated by 
non-overlapping but self-intersecting loops at a point is precisely
the free group factor with an infinite number of generators \cite{D,G,D1,V}, a type
$II_1$ non-hyper finite von Neumann algebra \cite{V}.
However functional techniques show that the master field, that solves the loop
equations of large-$N$ $QCD_2$ for closed self-avoiding loops, lives instead inside a commutative
hyper-finite von Neumann algebra and indeed it can be determined
by the distribution of the eigenvalues \cite{GW}.
This fact suggested to the author to look for hyper-finite
solutions of the loop equations for closed self-avoiding loops in the four
dimensional case.
In a sense this approach follows the original Witten proposal for the master
field \cite{W}, that contained implicitly some kind of finite dimensional approximation
property in the prescription of evaluating gauge invariants operators on the
master field by means of the normalised approximately finite dimensional matrix trace
$\lim _N \frac{1}{N} Tr_N$. In fact the existence of an approximately finite
dimensional trace is weaker than hyper-finiteness but implies that the
von Neumann algebra of the corresponding GNS representation is embeddable
in the factor $R^{\omega}$ \cite{Brown}, the ultra power of the unique hyper-finite
factor $R$ (see appendix A). \\
The plan of the paper is the following one.
Sect.(2) is an algebraic detour in which we recall some old and new
ideas about the master field. Though it is not strictly necessary for the rest
of the paper it helps to put sect.(3) and sect.(4) in the proper algebraic 
perspective. We make also some comment on the physical meaning of
restricting to closed self-avoiding loops.
In addition in sect.(2) we recall the quenched \cite{EK, Neu} and twisted
\cite{Twc} Eguchi-Kawai (EK) \cite{EK} reductions that are known to be equivalent
to the large-$N$ limit of $QCD$. The quenched and twisted theories that 
we refer to are the continuum version of their original lattice 
formulation \cite{EK,Twl1,Twl2}. 
A partially quenched \cite{RT} or twisted theory \cite{Twc}
is used in this paper as a technical tool to embed an inductive sequence of
finite dimensional Hitchin systems into the $QCD_4$ functional integral. \\
Sect.(3) contains a first change of variables, from the gauge connection to the
moduli fields of finite dimensional gauge orbits obtained by hyper-Kahler \cite{Hi, Hi2}
reduction in a partially quenched Eguchi-Kawai model or to
the moduli fields of the hyper-Kahler reduction of a certain infinite
dimensional central extension of Hitchin systems in the twisted case.
In this section the functional integral is represented as an integral over the
(energy) levels of the microcanonical ensemble for the action of the local gauge
group. However an effective action on the gauge orbits, that turn out to be
Hitchin systems (or an infinite dimensional central extension of, in the twisted
case), cannot be derived by the loop equations for these variables yet.\\
In sect.(4) the effective action is constructed by means of a further change of
variables, from the moduli fields of the orbits of the $SU(N)$ gauge group to
the moduli fields of the corresponding holomorphic (see appendix B) de Rham
local system (or a central extension of).\\
Objects related to the holomorphic de Rham local system have been introduced
under the name of opers by Beilinson and Drinfeld in their study of the
geometric Langlands correspondence \cite{BD}(opers will play a role in
sect.(6), after quantising Hitchin systems as an
alternative tool to solve the loop equations). 
The construction of the effective action involves a de Rham local system for a
partially quenched or twisted Eguchi-Kawai model. 
Hyper-finiteness is shown essentially by construction
in the quenched version, while it follows with little more effort
in the twisted theory, since in this case
the gauge algebra has to be defined from start on
an infinite dimensional Hilbert space and only later on approximated by
compressions that define finite dimensional representations of the fundamental
group. In any case hyper-finiteness should hold also in the twisted theory, at
least because of the equivalence of the quenched and twisted theories in the
large-$N$ limit. \\
In sect.(5) the effective action is computed in terms of functional determinants.
The effective action is the sum of three terms.
The first one is the classical action of the EK partially quenched or twisted model.
The second one is the logarithm of a functional determinant that arises
from localisation on the microcanonical ensemble.
The third one is the logarithm of an "anomalous" Jacobian that
arises from the change of variables from the moduli of Hitchin systems to the
moduli of the holomorphic (i.e. chiral) de Rham local systems. 
Despite the techniques employed in this paper are mostly based on two
dimensional differential geometry, we find that the localisation determinant
on the microcanonical ensemble of the Hitchin systems can be computed in a
manifestly four dimensional Euclidean (or Lorentz) invariant way.
Remarkably we find that the first coefficient of the beta function is
exactly reproduced, at lowest order in the expansion in powers of the moduli
fields (that at lowest order coincides with the usual perturbative expansion),
by the contribution of the localisation determinant in the master equation.
It follows immediately that the techniques employed here apply also
to the ${\cal N}=1$ super-symmetric extension of large-$N$ $QCD_4$ obtained adding 
to the pure gauge theory minimally coupled Majorana gluinos in the adjoint representation,
since in this case the ${\cal N}=1$ effective action is
obtained simply adding the logarithm of the fermion Pfaffian to the effective
action of the pure gauge theory. \\ 
In sect.(6), using ideas based on the geometric Langlands correspondence
\cite{BD,F,F1,F2,F3,F4,F5} (and references therein),
we suggest another way of solving loop equations for closed self-avoiding
loops by quantising Hitchin systems. In this last formulation,
due to quantisation, hyper-finiteness of the ambient algebra may get lost,
keeping nevertheless (quantum) integrability.\\
In this case the master field becomes an element of a quasi-conformal (vertex) algebra. This
is very close to interpreting the master field as an element of a
vertex algebra of a conformal field theory, as it would
be required by the conjectured gauge-field/string duality
\cite{Po1,Po2}, but for
the fact that our algebra is obtained first solving the Hitchin self-duality equations
on a punctured torus and then quantising the resulting moduli space, while
in the usual construction of conformal blocks first the space of two dimensional
connections is quantised and then the quantised
Hitchin equations are imposed as constraints on the space of physical states
\cite{Wea,Wea1}. \\
Our conclusions are contained in sect.(7). \\
In appendix A we recall some elementary facts about von Neumann algebras,
that may help to understand the revival of the algebraic approach to
the master field. \\
In appendix B we remind the reader of several notions of geometrical nature,
about Hitchin bundles, that are used in the paper. \\
In appendix C we compare our approach to the loop equations with Witten localisation
formulas in $QCD_2$ \cite{W2,DH,Bis1,Bis2} . \\
In appendix D we compute, in a manifestly Euclidean (or Lorentz) invariant
way, the divergences of the localisation determinant
at lowest order in the expansion in powers of the moduli fields.

\section{Loop equations: an algebraic detour.}

Recently there has been a revival of applications of ideas in the theory 
of von Neumann algebras (for a comprehensive review see \cite{C}) to field
theory, in particular to the large-$N$ limit (see for a recent review
\cite{Ma}).
We regard the large-$N$ limit as defined by the loop equations \cite{MM,MM1}, that
are the Schwinger-Dyson equations in the large-$N$ limit for
the expectation values of traces of Wilson loops in gauge theories or for a 
complete set of gauge invariant operators in other theories
(i.e. matrix models).
They are obtained noticing that the integral of a derivative vanishes:
\bea
 0  =  \int DA_{\mu} Tr \frac{\delta}{\delta A_{\nu}(z)}
  \exp(-\frac{N}{4 \lambda} \int Tr F_{\mu \nu}^2 d^4x) \Psi(x,x;A)= \nonumber \\
  =  \int DA_{\mu} \exp(-\frac{N}{4 \lambda} \int Tr F_{\mu \nu}^2 d^4x) 
  (\frac{N}{ \lambda}  Tr(D_{\mu} F_{\mu \nu}(z) \Psi(x,x;A))+ \nonumber \\
   +i\int_{C(x,x)} dy_{\nu} \delta^{(d)}(z-y) Tr(\lambda^a \Psi(x,y;A) \lambda^a 
  \Psi(y,x;A)) )
\eea  
where the sum over the index $a$ is understood.
Here $ \lambda^a $ are Hermitian generators of the Lie algebra and 
\bea
\Psi(x,y;A)=P \exp i\int_{C_{(x,y)}} A_{\mu} dx_{\mu}
\eea
$\Psi(x,x;A)$ is the monodromy matrix of the connection $A_{\mu}$
along the closed loop $C(x,x)$ based at the point $x$, i.e. $\Psi(x,x;A)$ is
the Wilson loop.
For the group $SU(N)$ using the identity:
\bea
\lambda^a_{\alpha \beta} \lambda^a_{\gamma \delta}=
\delta_{\alpha \delta} \delta_{\beta \gamma}-
\frac{1}{N} \delta_{\alpha \beta} \delta_{\gamma \delta}
\eea
we get:
\bea
0=\int DA_{\mu}\exp(-\frac{N}{4 \lambda} \int Tr F_{\mu \nu}^2 d^4x)
(\frac{N}{ \lambda}  Tr(D_{\mu} F_{\mu \nu}(z) \Psi(x,x;A))+ \nonumber \\
 +i\int_{C(x,x)} dy_{\nu} \delta^{(d)}(z-y) Tr( \Psi(x,y;A)) Tr(\Psi(y,x;A))+ \nonumber
 \\ 
  -\frac{i}{N} Tr(\Psi(x,y;A)\Psi(y,x;A)))
\eea 
where the last term may be disregarded in the large-$N$ limit.
The second term is the quantum contribution to loop equations, while
the first term is the classical one.
Thus, using the factorisation of gauge invariant operators in the large-$N$
limit and 
noticing that the expectation value can be combined with the matrix trace
to define a new generalised trace $\tau$,
our problem is to find a (unique) solution $A_{\mu}(x)$ to
\bea
0=\frac{1}{ \lambda}  \tau (D_{\mu} F_{\mu \nu}(z) \Psi(x,x;A)) + \nonumber \\
  +i\int_{C(x,x)} dy_{\nu} \delta^{(d)}(z-y) \tau( \Psi(x,y;A)) \tau(\Psi(y,x;A)) 
\eea
for every closed contour $C$, with values in a certain operator algebra with
normalised ($\tau(1)=1$) trace $\tau$ \cite{D,G,D1}. Such a solution is named the master field
\cite{W}.
The recent revival concerns the precise identification of the algebra and the
trace relevant for the large-$N$ limit of field theories \cite{D,G,D1}.
Operator algebras have already played some role in the early days of
constructive field theory, but they have been later
abandoned in favour of functional integral techniques as soon as it became
clear that the algebra of a single Euclidean scalar field is commutative.
In our case however the algebra of Euclidean matrix fields in the
large-$N$ limit is highly non commutative. \\
The recent interest in the interplay between the theory of von Neumann algebras
and the large-$N$ limit was driven on the mathematical side by Voiculescu use
\cite{V} of large-$N$ matrix models as a tool in the study of non hyper-finite
von Neumann algebras, in particular of the "elusive" free group factors that 
are left out by Connes \cite{C} algebraic classification of all hyper-finite
factors (see appendix A). On the physical side, after Singer suggestion
\cite{Si} that type $II_1$ representations of Voiculescu free group factors
are the relevant ones for the
study of the large-$N$ limit, several old and new results about the master 
field have been
interpreted in the light of the theory of von Neumann algebras \cite{D,G,D1}.
These investigations culminated with the precise identification
of the ambient algebra and the trace for all matrix models
in the large-$N$ limit, at least to all orders in the planar perturbation theory
\cite{D,G,D1}.
In this section we recall some old and recent ideas about the master field,
interpreting them from the more recent algebraic point of view (see
\cite{Ma} for a review).
Though this is possibly historically incorrect, it is useful for our
purposes.
After the discovery of large-$N$ factorisation \cite{Mig} of gauge invariant
local operators, Witten suggested that the large-$N$ limit could be understood as
localisation of the functional integral on a unique gauge orbit, the master
orbit, on which gauge invariant operators should be evaluated by means of the
usual normalised matrix trace in the large-$N$ limit \cite{W}.
In the retrospect we observe that the use of this matrix trace is tantamount to
postulating that the von Neumann algebra generated by the master field
is of type $II_1$ \cite{Si} and admits an approximately finite dimensional trace,
in particular is embeddable into $R^{\omega}$, the ultra product of the unique
hyper-finite factor $R$. If in addition the trace is
uniformly approximately finite dimensional the associated GNS representation is
hyper-finite (HF) (see appendix A) \cite{Brown}.
While the aim of this paper is to find observables in large-$N$ $QCD_4$ for
which the original Witten idea of a possibly hyper-finite master field can be
concretely realized, in general the hyper-finiteness condition turned out too much
restrictive, even in the case of the large-$N$ limit of a free scalar field 
\cite{Haa1}.
Haan \cite{Haa2} discovered that the von Neumann algebra
of the master field for a free scalar field was the Cuntz algebra with
an infinite number of generators (though he did not give it this name).
Haan found the following representation for the master field of a free
scalar field in the large-$N$ limit:
\bea
\phi(x)=\int a(p) \exp(ipx)+a(p)^* \exp(-ipx) dp
\eea
with the operators $a(p)$ and $a(p)^*$ satisfying:
\bea
a(p) a(q)^*= \delta^{(d)}(p-q) \nonumber \\
a(p) \mid \Omega \rangle =0
\eea
where $ \mid \Omega \rangle $ is the vacuum state.
In addition he found, on the basis of the large-$N$ Schwinger-Dyson equations,
that also in the interacting case of a scalar theory
with quartic interaction (but in a way that trivially extends to any interaction)
the ambient algebra of the master field was the vacuum (i.e. tracial) GNS 
representation of the Cuntz algebra.
The Cuntz algebra in general is generated by the following creation and annihilation
operators, that satisfy:
\bea
a_i  a_j^*= \delta_{i j} 1
\eea
and no other relation but the completeness condition:
\bea
\sum_i a_i ^* a_i = 1-P_{\Omega}
\eea
where $ P_{\Omega} $ is the orthogonal projector into the vacuum.
A tracial representation is obtained requiring:
\bea
a_i \mid \Omega \rangle =0
\eea
where $\mid \Omega \rangle $ is the vacuum. After Voiculescu work we know
that the algebra generated by the self-adjoint elements $a_i^* +a_i$
is a type $II_1$ algebra not hyper-finite for $i \ge 2$.
The Cuntz algebra with more than one self-adjoint generator in its tracial
representation is algebraically isomorphic to the free group factor with an
equal number of generators, that was known already to von Neumann
to be an example of a non-HF factor \cite{C}.
The occurrence of free group factors in the large-$N$ limit of field 
theories is implicit in the work of \cite{Cv} on the effective action
of the large-$N$ limit via the introduction of symbols that "neither
commute nor anti-commute" (an informal definition of a free
algebra) and in a way or another enters (implicitly) most of the 
subsequent developments (for example in the use of the loop
algebra \cite{Jev, Jaf}) until it has been made explicit in the work of 
\cite{D,G,D1} 
through the introduction of the Cuntz algebra
as the ambient algebra for the master field. \\
In particular in \cite{D,G,D1} it is proved, to every order in planar perturbation
theory, that the master algebra is a sub-algebra of the Cuntz algebra
generated by the following elements:
\bea
M_{\mu}=a^*_{\mu}+M_{\mu \nu}a_{\nu}+M_{\mu \nu \rho} a_{\nu} a_{\rho}+...
\eea
where $M_{...}$ are the planar correlation functions, so that the knowledge
of the master field is in fact equivalent to the knowledge of the solution
of the planar limit.
To summarise, we know that the master algebra of a large-$N$ free field 
is a non-HF Cuntz algebra and that the master algebra of an interacting
large-$N$ matrix model is a sub-algebra of the Cuntz algebra, probably
non-HF too.
However in some very special cases, such as the one-matrix model \cite{PAR}, 
traces of local operators can be reconstructed by of the distribution
of the eigenvalues, that generates an Abelian HF algebra (in fact in this case 
the master algebra is the commutative von Neumann algebra generated by $M_{\mu}$
for just one $ \mu $). In this case HF does not come as a surprise, since
a commutative $C^*$-algebra is of type $I$ and therefore HF (see Appendix
 A). We regard the lack of HF in the general case as the most disappointing,
making difficult any progress in finding exact solutions of the loop equations
(unless, perhaps, some kind of exact integrability is present).
In this paper we renounce to attempt a solution of the large-$N$ limit for all
observables, but rather we concentrate on those physically relevant observables
for which a HF solution may exist.
In this search we use as a guide the important example of large-$N$ $QCD_2$
that is explicitly solvable \cite{KK}.
For $QCD_2$ a gauge exists in which the theory is Gaussian, so that
the master algebra is generated by the Cuntz algebra in that gauge \cite{D,G,D1}.
In addition the algebra generated by Wilson loops based at a point is the free
group factor with an infinite number of generators, one for each simple
loop, i.e. non overlapping self-intersecting loop at a base point \cite{G}.
This can be proved noticing \cite{G} that the iterative solution of the Migdal-Makeenko
equations with respect to the number of self-intersections coincides with the
iterative equation associated to a Voiculescu free family of non-commutative
random variables \cite{V}, i.e.:
\bea
\langle (W(A_1)-\langle W(A_1)\rangle)...(W(A_n)-\langle W(A_n)\rangle) \rangle =0
\eea
This provides iterative equations for the expectation values of products
of $n$ random variables in terms of expectation values of a lower number of
variables. The entire family is thus determined by the expectation values
of self-avoiding loops.
For each simple loop of area $A_k$ it is possible to give an explicit
unitary master field in terms of the Cuntz algebra \cite{G}:
\bea
W(A_k)= \exp(i \sqrt A_k (a_k+a_k^*))
\eea
However the expectation value of a closed
self-avoiding Wilson loop on a large sphere 
can be computed by means of the distribution of the eigenvalues and as such
admits a master field that generates a commutative HF algebra, as the master field of
the one-matrix model. Indeed, on a sphere of large area, the average
of a Wilson loop of area $A$ is given by \cite{D2}:
\bea
W(A)=Z^{-1} \int \prod d\theta_i \prod_{i \ne j}(\theta_i-\theta_j)
\exp(-\frac{N}{2A} \sum_i \theta_i^2) \times \nonumber \\
\times \sum_i N^{-1} cos \theta_i
\eea
where:
\bea
Z= \int \prod d\theta_i \prod_{i \ne j}(\theta_i-\theta_j)
\exp(-\frac{N}{2A} \sum_i \theta_i^2) 
\eea
In the large-$N$ limit the distribution of the eigenvalues is determined by the
saddle-point equation:
\bea
\frac{N}{2A}  \theta_i=\sum_{j \ne i} \frac{1}{(\theta_i-\theta_j)}
\eea
In sect.(4) we will see that the existence of HF solutions of the
loop equations is somehow
linked to the condition of non self-intersection.
But even in this case, considerable difficulties arise 
when a "simple"
(i.e. not living in a free group algebra)
solution of the loop equations is attempted. Indeed for a closed 
self-avoiding loop based at $x$  the loop equations degenerate into:
\bea
0=\int DA_{\mu}\exp(-\frac{N}{4 \lambda} \int Tr F_{\mu \nu}^2 d^4x)
(\frac{N}{ \lambda}  Tr((D_{\mu} F_{\mu \nu}(x) + \nonumber \\
 + i\int_{C(x,x)} dy_{\nu} \delta^{(d)}(x-y) Tr(1)) \Psi(x,x;A))) 
\eea  
We see therefore that, despite the equations have been "linearised"
in such a way that the classical and quantum terms have the same
operator structure, the quantum contribution has still a different 
functional dependence on the loop $C$ from the classical term,
because the distribution:
\bea
i\int_{C(x,x)} dy_{\nu} \delta^{(d)}(x-y)
\eea
is obviously loop dependent.
This is even more manifest writing the loop equations for self-avoiding loops
in the Migdal-Makeenko form \cite{MM,MM1}:
\bea
0= \partial_{\mu}(x) \frac{\delta \langle Tr\Psi(x,x;C) \rangle
}{\delta \sigma_{\mu \nu}
(x)} +  \nonumber \\
 + \int_{C(x,x)} dy_{\nu} \delta^{(d)}(x-y) Tr(1) \langle Tr\Psi(x,x;C) \rangle 
\eea
Somehow anticipating the result of sect.(4) we will manage to change variables 
in the loop equations in such a way that in the new variables the corresponding
distribution in the quantum contribution will be loop independent for self-avoiding loops.
This will open the way to find an equivalent effective action. \\
Since we want to solve loop equations only for closed self-avoiding
loops we wonder how much physical information is lost
by this restriction. Obviously we can reconstruct by such loops the string tension
at large distances and the running of the coupling constant at short distances.
The next problem is the spectral information, i.e. the meson or the
glueball spectrum. This turns out to be dependent on the dimension of
space-time.
Essentially, for the meson spectrum, we want to compute the Dirac propagator in the background
of the master field. By employing a path integral representation for the
propagator, we represent it as an average over Wilson loops
computed along Brownian paths (for a recent use of this representation in the framework
of the large-$N$ limit of $QCD$ see \cite{Mig2,Mig3}).
Now in $d=2$ Brownian paths self-intersect any number of times
with probability one, while in $d=4$ they are self-avoiding with
probability one (see for example \cite{Simon}).
Therefore in $d=4$ the spectral information is contained in self-avoiding
loops. An analogous argument holds for the glueball spectrum.
We should mention that the construction of the effective action that we
present in this paper holds only
for planar loops, while in general to reconstruct the meson or the
glueball spectrum we would
need also non-planar ones. Since, however, every loop can be made to lie on a
surface, we think that our construction can be extended to the general case at
the price of making the geometry more involved, though this will not be
attempted in this paper.
We end this section recalling the EK reduction \cite{EK}, that plays a significant role
in our approach to the loop equations. 
The quenched EK reduction \cite{EK, Neu} is essentially a way of constructing a theory with
the same master field, i.e. the same large-$N$ loop equations, but
with fluctuations reduced by a factor of $N^{-1}$.
In the process space-time degrees of freedom are reduced to lower,
eventually zero dimensions.
The partition function of the quenched EK model is given by:
\bea
Z= \lim_{N \rightarrow \infty} \int dA_{\mu} \exp(\frac{N(2 \pi)^d}{4\lambda
\Lambda^d} 
Tr[p_{\mu}+A^{ch}_{\mu},p_{\nu}+A^{ch}_{\nu}]^2)  
\eea
where the integration is now over constant $N \times N $ matrices and
$p_{\mu}$ are diagonal matrices with uniform distribution of eigenvalues,
i.e. $p_{\mu}$ are the quenched momenta.
In the Eguchi-Kawai
quenched prescription the volume of the phase space divided by $(2 \pi)^d$
is assumed to scale as $N$ according to $ (2
\pi)^{-d} \Lambda^d V=N$
in such a way that the action is effectively rescaled by a factor of $N$
in order to compensate for a
reduction of a factor of $N$ in the entropy of the functional integration,
that now is only over constant matrices.
One way of seeing that the EK model is equivalent to the large-$N$ limit of
$QCD$ is to look at its loop equations:
\bea
0= \int dA_{\mu} \exp(\frac{N(2 \pi)^d}{4\lambda
\Lambda^d}  Tr[p_{\mu}+A_{\mu},p_{\nu}+A_{\nu}]^2)
\times \nonumber \\  
\times (\frac{N}{\lambda} Tr[p_{\mu}+A^{ch}_{\mu},[p_{\mu}+A^{ch}_{\mu},p_{\nu}+A_{\nu}]]
\Psi(x,x;A)+ \nonumber \\
-i\int_{C(x,x)} dy_{\nu}\frac{\Lambda^d}{(2 \pi)^d} Tr( \Psi(x,y;A)) Tr(\Psi(y,x;A))) 
\eea
Provided that the coupling
constant was rescaled as in Eq.(20), these are the same loop equations as for large-$N$ continuum
$QCD$, but for the fact that there is no
delta function in the quantum contribution, since in the reduced model
we are integrating over constant matrices. Therefore the loop equations of the
reduced EK model coincide with the ones of the un-reduced theory provided the
trace of open loop vanishes, that is a way of reproducing the 
delta function. This is ensured by the global $R^d$
symmetry (if it remains unbroken) (see \cite{Ma}):
\bea
A_{\mu} \rightarrow p_{\mu}1+A_{\mu}
\eea
that implies:
\bea
\frac{1}{N} \langle Tr(\Psi(x,y;A)) \rangle = \exp(ip(x-y))  \langle 
\frac{1}{N} Tr(\Psi(x,y;A)) \rangle
\eea
In fact the quenching prescription was introduced precisely for this aim.
A variant of the quenched EK model is the twisted model \cite{Twc}:
\bea
Z= \int dA_{\mu} \exp(\frac{N}{4\lambda} Tr[\Gamma_{\mu}+A_{\mu},\Gamma_{\nu}+A_{\nu}]^2)  
\eea
where $\Gamma_{\mu}$ are (unbounded) self-adjoint operators acting
on an infinite dimensional Hilbert space satisfying: 
\bea
[\Gamma_{\mu},\Gamma_{\nu}]= i(\theta^{-1})_{\mu \nu}1
\eea
with $(\theta^{-1})_{\mu \nu}$ vanishing in the large-$N$ limit.
In modern language the twisted model is a gauge theory
defined over the non-commutative Euclidean space:
\bea
[x_{\mu},x_{\nu}]= i \theta_{\mu \nu}1
\eea
The non-commutative Yang-Mills theory (for a review see \cite{Ne}) equivalent to the twisted model
is obtained working with commuting coordinates and replacing
the ordinary product between functions with the Moyal star product
defined by:
\bea
f(x) \star g(x)= \exp(i\theta_{\mu \nu}
\frac {\partial}{\partial y_{\mu}} \frac {\partial}{\partial z_{\nu}})
f(x+y)g(x+z)\mid_{y=z=0}
\eea
Thus the partition function of the twisted theory can be re-casted into its
non-commutative form, that looks like a matrix model:
\bea
Z= \int dA_{\mu} \exp(-\frac{N}{4\lambda} 
Tr(i[C_{\mu},C_{\nu}]-(\theta^{-1})_{\mu \nu}
)^2)  
\eea
where $C_{\mu}=(-i \theta^{-1})_{\mu \nu} x_{\nu}+iA_{\mu}$ and the
product is the Moyal product.
The loop equations of the non-commutative theory \cite{loop1,loop2} differ from those of the
commutative theory for the need of taking an integral over the base-point
of the Wilson loop, in order to get gauge invariant quantities,
since the trace
now includes integration over the base point of the Wilson loop.
However, after taking the expectation value,
the trace of Wilson loops is proportional to the volume of space-time,
because of translational invariance,
so that we get a normalised trace dividing by the volume.
In the $\theta_{\mu \nu} \rightarrow \infty$ limit the loop equations
of the non-commutative theory (for any positive $N$, even $N=1$)
coincide with the ones of the large-$N$ commutative theory.

\section{The microcanonical ensemble.}

It is quite well known that every Euclidean field theory can be thought
as the partition function of a system of classical statistical mechanics
in which the coupling constant squared is identified with
the classical temperature (see for example \cite{Pol3}).
Following this interpretation the Euclidean functional integral:
\bea
\int \exp(-\frac{1}{\lambda} S(A)) DA
\eea 
looks like the classical partition function:
\bea
\int \exp(-\beta H(p,q)) dp dq .
\eea
The second integral has an obvious representation as an integral over the
energy levels of the microcanonical ensemble (see for example \cite{Ruelle}):
\bea
\int \exp(-\beta H(p,q)) dp dq =\nonumber \\
= \int \exp(-\beta E) \delta(H(p,q)-E) dp dq dE = \nonumber \\
= \int \exp(-\beta E) Det(\nabla H)^{-1} dE 
\eea
The aim of this section is to find an analogous representation for the partition
function of $QCD_4$. The motivation is the following one. The original Witten
proposal for the master field can be interpreted as a sort of localisation
of the functional integral at large-$N$ on a certain microcanonical ensemble
specified by the set of parameters that describe the master gauge orbit.
It is therefore natural to classify explicitly gauge orbits.
According to the discussion in sect.(2) we assume that the relevant gauge
orbits are those of hyper-finite type, i.e. those which can be obtained as limits
of finite dimensional orbits.
In sect.(4) we will inquire whether and to what extent localisation on this
microcanonical ensemble is consistent with the loop equations.
In this section we simply define the microcanonical ensemble.
Following our microcanonical
analogy it is clear that the role of the energy
should be played by a set of levels of the Hamiltonians that parameterise the 
gauge orbits. The levels will be therefore part of the moduli of the
orbits.
In addition, since the the gauge group acts locally, the moduli will be moduli
fields. 
The representation that we are going to find
is an identity as it is Eq.(31) and there is a priori no guarantee that
it will help us to derive an 
effective action defined over the microcanonical ensemble
from the loop equations. Indeed it is still
necessary to functionally integrate over 
the moduli fields, whose fluctuations a priori will contribute loop dependent
quantum terms in the loop equations, as it happens for the original integration
variables, i.e. the gauge connection.
In the next section we will show that an effective action exists 
for (planar) closed self-avoiding loops, provided we make a further change of
variables from the levels of the Hamiltonians to the moduli of the associated
holomorphic de Rham local system and 
we allow a central extension in the defining equations of the microcanonical
ensemble. The existence of the central extension
does not spoil the hyper-finiteness property in the case of the partially
quenched Eguchi-Kawai model, but HF is less obvious in the twisted theory
since in this case gauge orbits have to be infinite dimensional from
start. However the HF property should be kept also in this case,
since the quenched and twisted models are equivalent in the
large-$N$ limit, though we will not provide any direct proof of it.
Following our statistical-mechanical analogy we want to think the Euclidean
functional integral as an integral over the phase space rather than the
configuration space. $DA$ should be therefore a Liouville measure associated
to some symplectic structure. Since our functional integral has a four
dimensional nature it turns out that the Liouville measure $DA$ is the volume
form associated to three different symplectic forms: $\omega_K$ and 
$\omega_C$ ($\omega_C$ is complex,
so that its real and imaginary parts provide two real symplectic forms):
\bea
\omega_K=\int d^4x Tr(\delta A_z  \wedge \delta A_{\bar z} + \delta A_u \wedge 
\delta A_{\bar u})
\nonumber \\
\omega_C=\int d^4x Tr(\delta A_z \wedge \delta A_{\bar u}) 
\eea
In fact the space of four dimensional connections can be thought of as an 
infinite dimensional affine hyper-Kahler manifold.
For technical reasons that will be evident in sect.(4) we need a reduction of the
structure (gauge) group from four to two dimensions.
This is not restrictive in the large-$N$ limit since it is known by Eguchi-Kawai
that the theory can be reduced by quenching or twisting even to zero
dimension keeping the same large-$N$ limit.
After the partial quenching (twisting) mentioned before, two components of the original
connection still transform as a connection under the two dimensional structure
group while the other two components transform as a Higgs field.
We notice that the identification of two components of the gauge connection
with the Higgs field is not canonical, i.e. coordinate independent.
However in the case of $QCD_4$ we have a preferred choice of coordinates
that corresponds to a flat metric on $T^2 \times T^2$ or better to a conformally
flat metric, since the classical theory is conformally invariant in four
dimensions.
Our symplectic space can be now identified (non-canonically, i.e. in a
coordinate dependent way) with the cotangent bundle of two dimensional
connections, whose coordinates are the pairs $(A=A_z dz+A_{\bar z} d\bar z,
\Psi= \Psi_z dz+\Psi_{\bar z} d\bar z )$. 
More precisely in the partially quenched theory we have the identifications:
\bea
\Psi_z=-iD_u= -i(\partial_u + iA_u)=p_u+A^{ch}_u \nonumber \\
\Psi_{\bar z}=i \bar{D_ u}= i(\partial_{\bar u}-i A_{\bar u})=p_{\bar u}+A^{ch}_{\bar u} 
\eea
while in the partially twisted theory we have: 
\bea
\Psi_z=\Gamma_u+A_u \nonumber \\
\Psi_{\bar z}=\Gamma_{\bar u}+A_{\bar u} 
\eea
with the other two components of the gauge connection remaining the same
in both cases. Let us notice that we have used the quenching prescription to identify
covariant derivatives with the Higgs field in the quenched theory and
the twisting prescription for the analogous identification in the twisted
theory.
On the cotangent bundle the gauge transformations act by Hamiltonian vector
fields with
respect to each of the symplectic structures \cite{Hi,Hi2}.
The gauge orbits of the hyper-Kahler reduction are classified by the set of 
levels of the moment maps
(the Hamiltonians) for the action of the gauge group, one for each of the
three symplectic structures:
\bea
F_A-i\Psi^2=\mu^0 \nonumber \\
\bar{\partial_A} \psi=n \nonumber \\
\partial_A \bar{\psi}=\bar{n} \nonumber \\
\eea
with $\mu^0$ Hermitian and $n$ with values in $sl(N)$.
Therefore the partially quenched EK model admits a stratification 
by the following microcanonical ensemble:
\bea
Z=\int DA D\Psi D\mu^0 Dn D\bar{n} \exp(-S_{EK})  \times \nonumber \\
\times \delta(F_A-i\Psi^2-\mu^0) 
\delta(\bar{\partial_A} \psi-n)
\delta(\partial_A \bar{\psi}-\bar{n}) 
\eea
where the obvious resolution of identity has been inserted.
This equation has a meaning for every finite $N$ in the quenched
EK model, since the identification in Eq.(33) has a meaning at the level
of finite matrices.
For the twisted model the preceding equations have to be modified
because of the non-trivial commutator:
\bea
[\Gamma_u,\Gamma_{\bar u}]= \theta^{-1}1
\eea
that requires in addition that the gauge orbits be infinite dimensional
even before taking the large-$N$ limit.
In this case our microcanonical ensemble is given by:
\bea
F_A-i\Psi^2=\mu^0+ \theta^{-1}1\nonumber \\
\bar{\partial_A} \psi=n \nonumber \\
\partial_A \bar{\psi}=\bar{n} \nonumber \\
\eea
modulo gauge transformations, so that a central extension has been introduced
in the Hermitian moment map.
From Eq.(38) the microcanonical stratification for the partially twisted
theory follows: 
\bea
Z=\int DA D\Psi D\mu^0 Dn D\bar{n} \exp(-S_{TW})  \times \nonumber \\
\times \delta(F_A-i\Psi^2-\mu^0-\theta^{-1}1) 
\delta(\bar{\partial_A} \psi-n)
\delta(\partial_A \bar{\psi}-\bar{n}) 
\eea
The symplectic reduction in Eq.(35,38) and the associated stratification of the
functional integral are somehow formal, unless the structure of the moduli fields
is specified. In general, if we require that the symplectic reduction be
associated to a good moduli problem, i.e. separable, the moduli fields 
must be taken as a linear combination of delta functions, in order that the corresponding
flatness condition defines a (possibly projective)
representation of the fundamental group of the punctured torus (see appendix B).
Despite this restriction, these moduli fields are dense in the sense of
the distributions in the functional integral, that is all that we need to
define it. \\
In sect.(4) we will see that the existence of punctures,
i.e. of a boundary in our space-time torus, has a more fundamental meaning to
solve loop equations than it seems from the definition of the microcanonical
ensemble given here. 
We would like to terminate this section writing the resolution of identity
associated to the microcanonical ensemble in a way that makes manifest its four 
dimensional Euclidean (or Lorentz) covariant nature:
\bea
Z=\int DA_{\mu} D\mu^-_{\mu \nu} \exp(-S_{YM})  
 \delta(F^-_{\mu \nu}(A)-\mu^-_{\mu \nu}-(\theta^{-1})^-_{\mu \nu}1) 
\eea
where the $^-$ superscript refers to the projection into the anti-selfdual part
of an antisymmetric tensor.
In Eq.(40) we have written the resolution of identity of the twisted
model, but the one of the quenched model can be obtained simply setting
$ (\theta^{-1})^-_{\mu \nu}=0$ and interpreting covariant derivatives
appropriately. In sect.(5) we will make use of this representation
to get an explicit Euclidean (or Lorentz) invariant form of the localisation
integral.

\section{Solving loop equations by Hitchin systems.}

In this section we consider loop equations for the connection $B=A+i\Psi$
in the partially quenched or twisted model. The advantage of using $B$ as
opposed to $A_{\mu}$ resides in the fact that planar loops for $B$ generate
a complete set of gauge invariant observables. The disvantage consists in the
fact that $B$ is not Hermitian so that $\Psi(x,x;B)$ is not unitary in general.
We will see later on under which conditions planar unitary monodromies, i.e.
$\Psi(x,x;A)$,
can be obtained directly from $\Psi(x,x;B)$.  
The loop equations for $B$ read:
\bea
0=\int DB_{\alpha} \exp(- S_{YM})
(Tr(\frac{\delta S_{YM}}{\delta B_{\alpha}(z)} \Psi(x,x;B))+ \nonumber \\
 - i\int_{C(x,x)} dy_{\alpha} \delta^{(2)}(z-y) Tr( \Psi(x,y;B)) Tr(\Psi(y,x;B)))
\eea 
with $\alpha=1,2$,
so that in these variables they are as difficult to solve as for the original
four dimensional connection $A_{\mu}$. The quantum contribution in general is
loop dependent and this dependence is an obstruction to the existence of a
conventional (hyper-finite?) effective action (we take the point of view that
an effective action defined directly in loop space \cite{Jev,Jaf}, from which the loop equations
may be derived as critical equations, probably would not be hyper-finite).\\
After implementing in the functional integral the resolution of identity by
means of the gauge orbits of the microcanonical ensemble of sect.(3),
we change variables in the loop equations from Hitchin bundles
to the corresponding holomorphic de Rham local system 
(see appendix B). In \cite{MB} an analogous change of variables, from the Hitchin
system to the action-angle variables of the Hitchin fibration of the
corresponding Dolbeault moduli space (see appendix B), was proposed on the basis of
energy/entropy considerations. \\ 
Our basic philosophy consists in changing variables in such a way that
the quantum contribution may be computed explicitly by evaluating a residue
that depends only on the topological class of the loop.
It is well known by the Cauchy formula that the line integral along
a closed loop of a holomorphic
function times the Cauchy kernel with pole not lying on the loop depends
only on the winding number of the loop around the pole
and on the value of the holomorphic function at the pole of the Cauchy kernel:
\bea
f(z) Ind_{C}(z)= \frac{1}{2 \pi i} \int_C \frac{f(w)}{z-w}dw
\eea
where $Ind_{C}(z)$ is the winding number of $C$ around $z$.
The existence would become possible of an effective action, for loops
in a certain topological class determined by the winding number, if only 
the Cauchy kernel could be generated by functionally differentiating 
the monodromy of the connection in the loop equations with respect to the
integration variables in the functional integral. 
We have seen in sect.(2) that functional differentiation with respect to
the gauge connection produces a delta function in the quantum contribution
to the loop equations.
The change of variables to the de Rham systems is thought in such a way that functional
differentiation of the monodromy with respect to the moduli fields of 
the holomorphic de Rham local systems
produces precisely the Cauchy kernel in the loop equations.
However, even if we reduce the computation of the quantum contribution
to the evaluation of a line integral of a Cauchy kernel, there are two
obstructions for the residue theorem to apply. The first one is that the
monodromy is not a holomorphic function of its endpoints. The second one is that
gauge invariance requires that the functional derivative with respect to the integration
variable be taken at a point
of the loop, for the result to be non-vanishing, i.e at a singularity
point of the Cauchy kernel.  
We will see below how each of these obstructions is the cure for the other
one.
After introducing the non-Hermitian connection $B=A+i\Psi=b+\bar b$ the 
microcanonical representation of sect.(3) holds in the form:
\bea
Z=\int 
\delta(F_B-\mu) 
\delta(d^*_A \Psi- \nu) 
\exp(-S_{EK}) DB D \mu D \nu 
\eea
where we have re-casted the microcanonical resolution of identity in a different,
equivalent form (see appendix B), for later convenience, in order to compute
the effective action (sect.(5)).
We now change variables from the Hitchin bundles to the holomorphic de Rham
local systems (see appendix B), whose moduli are described by the field $\mu'$,
obtained from the equation:
\bea 
F_B-\mu=0 
\eea
by means of a complexified gauge transformation $G(x;B)$ that puts the connection
$B=b+\bar b$ in the holomorphic gauge $\bar b=0$:
\bea 
\bar{\partial}b-i\mu'=0 
\eea
where $\mu=\mu^0+(n-\bar n)$ and $\mu'=G \mu G^{-1}$.
The partition function in terms of the moduli fields of the de Rham local
system becomes:
\bea
Z=\int 
\delta(F_B-\mu) 
\delta(d^*_A \Psi-\nu) 
 \exp(-S_{EK}) DB  \frac{D (\mu ,\nu)}{D(\mu',\bar{\mu}')} 
D \mu'  D \bar{\mu}'
\eea
The integral over $B$ can now be performed and the resulting functional
determinants, together with the Jacobian of the change of variables to the
de Rham local system, absorbed into the definition of the effective action $\Gamma$.
$\Gamma$ will be written more explicitly in the next section in terms
of these functional determinants. Let us notice that for computing functional determinants
a choice of metric is necessary. The natural metric for us is the
translational invariant metric on the torus $ \exp(-\rho) dz d\bar{z}$ with the
conformal factor constant. However the classical action, being conformally
invariant, does not depend on the choice of $\rho$. Yet we will see
below that our regularization of the quantum contribution to the loop equations
will require to transform the metric by a conformal diffeomorphism.
The partition function is now:
\bea
Z=\int \exp(-\Gamma) D\mu' D\bar{\mu}'
\eea
To realize our aim of getting the Cauchy kernel by functional differentiation
with respect to the moduli fields of the de Rham local system it is convenient
to study the loop equations for the Wilson loop constructed by the connection
$b$, thought as a functional of $B$ corresponding to gauge transforming $B$ into
the gauge $\bar b=0$. Such a gauge transformation belongs to the
complexification of the gauge group and it is rather a change of variables than
a proper gauge transformation. However, because of the property of the trace,
for closed loops, it preserves the trace of the monodromy. This will allow us,
later in the course of our computations, to transform the loop equations thus
obtained into equations for the monodromy of $B$. In our derivation of the loop 
equations, a crucial role is played by the condition that the expectation value
of an open loop vanishes. For open loops constructed by means of the connection
$B$ this property follows directly from the local gauge invariance of the
partially quenched or twisted EK model. However, the starting point in our case
is the monodromy of the connection $b$, whose expectation value is also
required to vanish for open loops.
Yet, this condition does not follow from gauge invariance, since the change of
variables to the de Rham local system ($\bar b=0$) does not allow any residual
local compact gauge freedom (unless perhaps we analytically continue to Minkowski space-time,
see below).
We will present here two slightly different derivations of the effective action
each based on a different way of achieving the vanishing of the expectation
value of open $b$ loops. \\
The most direct procedure consists in observing that in the partially quenched
theory the vanishing of the expectation value of open $b$ loops follows from the fact that the partial
quenching is thought precisely to keep unbroken the $R^2$
symmetry $B \rightarrow B+i(p_u dz+ p_{\bar u} d\bar z)$ and that in the gauge
$ \bar b=0$ the $R^2$ symmetry is broken down to $b \rightarrow b+i p_u dz $.
If this residual symmetry is unbroken, then the expectation value of the trace 
of open $b$ loops vanishes because of the equation:
\bea
 N^{-1}\langle Tr \Psi( z_1,z_2;b) \rangle =
 \exp(p_u( z_1- z_2)_z)
 N^{-1}\langle Tr \Psi(z_1,z_2;b) \rangle 
\eea
provided the normalised trace stays uniformly bounded in the large-$N$ limit.
To be precise, this holds only for loops that are open once projected on the
$z$ direction, because the $R^2$ symmetry is partially broken to
$b \rightarrow b+i p_u dz $. However this weaker property is all that is needed
in our loop equations.
In the other proof of the loop equations that we propose below, the assumption 
of uniform boundness is not necessary and the vanishing of the expectation
values of open $b$ loops holds without further specifications.
We may thus derive our loop equations in the partially quenched theory:
\bea
0=\int  D\mu' D\bar{\mu}'  Tr \frac{\delta}{\delta \mu'(w)}
  (\exp(- \Gamma)
  \Psi(x,x;b))= \nonumber \\
 = \int  D\mu' D\bar{\mu}' \exp(-\Gamma)
  (Tr(\frac{\delta \Gamma}{\delta \mu'(w)} \Psi(x,x;b))+ 
  \nonumber \\ 
  +\int_{C(x,x)} dy_z \bar{\partial}^{-1}(w-y) 
  Tr(\lambda^a \Psi(x,y;b)
  \lambda^a \Psi(y,x;b)) )= \nonumber \\
 =\int D\mu' D\bar{\mu}' \exp(-\Gamma)
  (Tr(\frac{\delta \Gamma}{\delta \mu'(w)} \Psi(x,x;b))+ \nonumber \\
  + \int_{C(x,x)} dy_z \bar{\partial}^{-1}(w-y)Tr( \Psi(x,y;b))
    Tr(\Psi(y,x;b))+ \nonumber \\
  - \frac{1}{N} Tr( \Psi(x,y;b) \Psi(y,x;b)))
\eea
that in the large-$N$ limit reduce to:
\bea
0=\int D\mu' D\bar{\mu}'  \exp(-\Gamma)
  (Tr(\frac{\delta \Gamma}{\delta \mu'(w)} \Psi(x,x;b))+ \nonumber \\
  + \int_{C(x,x)} dy_z \bar{\partial}^{-1}(w-y)Tr( \Psi(x,y;b))
    Tr(\Psi(y,x;b))) 
\eea
The most interesting case is when $w$ lies on the loop $C$, because
the loop equations can be transformed easily into an equation for $B$.
However in this case the contour integration includes the pole of the Cauchy
kernel. We need therefore a gauge invariant regularization.
Before doing so it is convenient to make a conformal diffeomorphism to put
the contour integral in standard form. We wish to do that because, when we
regularize the theory, we require a loop-independent universal
regularization. Hence it is natural to use the uniformization theory
of Riemann surfaces with a boundary.
It is not restrictive to take $w$ as the base point of the loop.
For every simple loop (i.e. non-overlapping, self-intersecting only at the
base point) we make a conformal transformation that maps each
petal into a disk and then each disk into the upper half-plane.
Each petal of the loop is then mapped into a copy of 
the upper half plane and the boundary identified with the $x$-axis.
Let $t$ be the local coordinate on the upper half-plane corresponding to this
conformal transformation. Obviously $t$ depends on $C$ and the microcanonical
resolution of the identity in the functional integral, once written in term of 
$t$, depends on $C$ too.  The loop equations in the variables associated to 
the coordinate $t$ look like the ones in the original global coordinate on the
torus $z$. Therefore we get:
\bea
0=\int D\mu'_{t \bar t} D \bar{\mu}'_{t \bar t} \exp(-\Gamma)
  ( Tr(\frac{\delta \Gamma}{\delta \mu'_{t \bar t}(w)} \Psi(x,x;b))+ \nonumber \\
  + \int_{C(x,x)} dy_t \bar{\partial}^{-1}(w-y)Tr( \Psi(x,y;b))
    Tr(\Psi(y,x;b))) 
\eea
with the important difference that the contour integral is now restricted to
the real axis and that the metric needed to define the functional determinants 
in $\Gamma$ is the conformally transformed metric $ \exp(-\rho) dt \bar{dt}
=\exp(-\rho) |\frac{\partial t}{\partial z}|^2
dz d\bar{z}$.
In $d=4$ it is expected that this conformal transformation will affect the effective
action only through its dependence on the conformal anomaly \cite{conf}.
A natural regularization of the quantum contribution to the loop equations
is a $i \epsilon$ regularization of the Cauchy kernel. This is the most 
natural among any other possible regularization in Euclidean space,
since it is in fact compatible with the analytic continuation to Minkowski
space-time, that manifestly preserves gauge invariance. The idea that Feynmann diagrams in two dimensions could
be regularized by analytical continuation to Minkowski space-time in certain cases
is contained in \cite{Pol3}.
A master field living
in the Cuntz algebra in Minkowski space-time has been
constructed in general, and in particular for Yang-Mills theories, in 
\cite{Arefeva}.
The result of the $i \epsilon$ regularization is the sum of
two distributions, the principal part plus
a one dimensional delta function: 
\bea
\bar{\partial}^{-1}(w_x -y_x +i\epsilon)= (2 \pi)^{-1} (P(w_x -y_x)^{-1}
- i \pi \delta(w_x -y_x))
\eea
The loop equations thus regularized look like:
\bea
 0=\int D\mu' D \bar{\mu'} \exp(-\Gamma) 
  (Tr(\frac{\delta \Gamma}{\delta \mu'(w)} \Psi(x,x;b))+ \nonumber \\
   + \int_{C(x,x)} dy_x(2 \pi )^{-1} (P(w_x -y_x)^{-1}
  - i \pi \delta(w_x -y_x)) \times \nonumber \\
  \times Tr( \Psi(x,y;b)) Tr(\Psi(y,x;b)))
\eea
where we are omitting the $t$ under-script.
Being supported on open loops the principal part does not contribute and the
loop equations reduce to:
\bea
0=\int D\mu' D\bar{\mu}' \exp(-\Gamma)
  (Tr(\frac{\delta \Gamma}{\delta \mu'(w)} \Psi(x,x;b))+ \nonumber \\
   -\int_{C(x,x)} dy_{x}
\frac {i}{2}\delta(w_x -y_x) Tr( \Psi(x,y;b)) Tr(\Psi(y,x;b)))
\eea
Taking $w=x$ and using the gauge transformation properties of the $b$ monodromy
and of $\mu(x)'$, the preceding equation can be rewritten in terms
of the connection $B$ and the moduli fields $\mu$:
\bea
0=\int D\mu' D\bar{\mu}' \exp(-\Gamma)
  (Tr(\frac{\delta \Gamma}{\delta \mu(x)} \Psi(x,x;B))+\nonumber \\
   -\int_{C(x,x)} dy_{x}
\frac {i}{2}\delta(x_x -y_x) Tr( \Psi(x,y;B)) Tr(\Psi(y,x;B)))
\eea
where we have used the condition that the trace of open loops vanishes
to substitute the $b$ monodromy with the $B$ monodromy.
This is our final form of the regularized Euclidean loop equations.
Let us notice that there is a sign ambiguity in the quantum contribution
that depends on the choice of the sign of $\epsilon$ in the
regulator or as to whether the conformal transformation is orientation
preserving or orientation inverting. 
The occurrence of the factor of $\frac{1}{2}$ seems to be related to the fact
that the pole of the Cauchy kernel lies along the contour integral. If the
contour
integral were to include the pole inside the loop and the residue theorem to apply
naively, the residue would be $1$, while it would be $0$ if the pole were outside the loop.
 Both choices however would violate
gauge invariance.  \\
For general self-intersection the Euclidean loop equations reduce to
(assuming the conformal map to be orientation preserving):
\bea
0=\int D\mu' D\bar{\mu}' \exp(-\Gamma)
  (Tr(\frac{\delta \Gamma}{\delta \mu(x)} \Psi(x,x;B))+ \nonumber \\
  -\frac{i}{2}\sum_{(y;y=x)} Ind_C(y) Tr( \Psi(x,y;B)) Tr(\Psi(y,x;B)))
\eea
Because of the product of traces, in general the quantum contribution does
not have the
same operator structure as does the classical term and it is loop dependent.
Therefore, in general, an effective
action need not to exist as a functional over the microcanonical
ensemble constructed by the inductive limit of finite dimensional orbits.
However, for self-avoiding loops, one of the traces in the quantum
contribution is just $Tr(1)=N$ and there is only one term as for the classical
contribution. 
It is now clear the role played by the absence of self-intersection for these
variables associated to the de Rham system of the microcanonical ensemble.
It is only in this case that the quantum contribution
is proportional to the monodromy as is the classical term.
In general we get a loop dependent factorized trace of monodromies, that cannot
be re-absorbed into a redefinition of $\Gamma$. \\ 
The result is the following Euclidean master equation for self-avoiding loops:
\bea
0=\int D\mu' D\bar{\mu}' \exp(-\Gamma) 
   (Tr((\frac{\delta \Gamma}{\delta \mu(x)}  
  - \frac{i}{2} Ind_C(x) Tr(1)) \Psi(x,x;B)))
\eea
that reduces to:
\bea
0=\frac{\delta \Gamma}{\delta \mu(x)}-\frac{i}{2} Ind_C(x) Tr(1)
\eea
The corresponding effective action is:
\bea
\Gamma_{eff}=\Gamma-\epsilon(C) \frac{i}{2} \int Tr \mu d^2x
\eea
with $\epsilon(C)=+1$ or $-1$. The two possible signs are due to the two
possible orientations of the loop $C$. In addition there is a sign ambiguity due to the conventional
choices that we made: the choice of regularization,
i.e $i\epsilon$ versus $-i\epsilon$ and the convention that
the conformal map is orientation preserving. 
Remarkably the sign ambiguity in the Euclidean loop equations corresponds
in fact to physical equivalence.
Indeed the quantum
contribution is a $\theta$-term with $\theta=\frac{1}{2}$. Now,
since $\int Tr \mu d^2x$ represents the first Chern class, $\theta$ is defined
modulo integers, so that for $\theta=\frac{1}{2}$ a change of sign does not
change $\theta$ modulo $1$. Had we regularized by analytical continuation to
Minkowski space-time we would have obtained the same imaginary quantum
contribution to the effective action, while the action $\Gamma$ would have been Wick
rotated by a factor of $i$.
In any case we see that the quantum contribution
for the de Rham system has produced a central extension in the master equation.
This means in general, that even if we start with a microcanonical ensemble
associated to $sl(N)$ bundles, central directions
in the moduli fields have to be included, otherwise the master equation cannot
be satisfied. Hence, in the partially quenched model
we have to extend our microcanonical ensemble 
from $sl(N)$ to $gl(N)$ and to allow a central extension in the Hermitian moment map as follows:
\bea
F_A-i\Psi^2=\mu^0 +c 1 \nonumber \\
\bar{\partial_A} \psi=n \nonumber \\
\partial_A \bar{\psi}=\bar{n}
\eea
After that the entire chain of derivations of this section can be consistently
performed to arrive at the master equation.
Yet, the occurrence of the $\theta$-term in the effective action is not completely
harmless, since it affects
the topological sectors of the $U(N)$ theory (see for example \cite{Alvarez}).
In fact, we will see in sect.(5)
that the central term in $\Gamma_{eff}$ is cancelled by a counter-term,
that arises as a $u(1)$-anomaly in the Jacobian of the "chiral" change of
variables from the Hitchin to the holomorphic de Rham systems, after a finite
conformal rescaling of the metric used to compute the functional determinants.
The "holographic" nature of our change of variables will be then completely
manifest, in the sense that we are cancelling the quantum boundary contribution
against one from the bulk, to obtain an effective action defined purely on the
bulk.\\
Summarising, given a sequence of solutions of the master equation for any finite $N$,
we have a sequence of solutions of the loop equations for finite $N$, up to terms
vanishing in the large-$N$ limit, corresponding to the omitted
connected contribution. Therefore for $N$ large but finite we have an
approximation of the
solution of the $N=\infty$ bare loop equations for planar self-avoiding loops
with accuracy $N^{-1}$.
The algebra generated by the monodromy of the master connection of
the partially quenched EK model
for any finite $N$ is a matrix algebra, since the microcanonical ensemble
defines a finite dimensional representation (of a central
extension) of the fundamental group of a punctured torus.
Obviously renormalization is necessary,
since we have written the master equation for the bare
master field, but in principle nothing forbids to perform it.
Assuming convergence of the renormalised master monodromy as $N$ goes to
infinity, it follows the hyper-finiteness of the master algebra
for planar closed self-avoiding $B$ loops. \\
It is also clear that the twisted model leads 
to the same master equation, with the only difference that now
the central extension is present from start and it is associated to an
infinite dimensional projective representation of the fundamental group. \\
To show hyper-finiteness also in the twisted case we must associate to this
infinite dimensional representation a system of approximating finite dimensional
representations. In the twisted case it is not restrictive to start with a
microcanonical ensemble that, though is infinite dimensional because of the
central extension due to the non-trivial commutator:
\bea
[\Gamma_{u},\Gamma_{\bar u}]= \theta^{-1} 1
\eea
has finite rank residues at the punctures. This implies that the monodromies
at the punctures are a direct sum of finite dimensional matrices and a multiple
of identity. The representation obtained projecting away the multiple of
identity part is the desired approximating system (finite dimensional approximations of
infinite dimensional representations of the fundamental group have been previously
considered in \cite{Matone} in the framework of non-commutative Riemann surfaces
and in the context of \cite{MB}).
To be explicit the master algebra is the algebra generated by an inductive
sequence of finite dimensional group algebras of representations of the
fundamental
group, where the inductive embedding is with respect to a partial order
on the growing divisor, D, and on the increasing rank, $N$, of the local system.
Thus we get in the quenched case:
\bea
\Psi(a) \Psi(b) \Psi(a)^{-1} \Psi(b)^{-1} \prod_p \Psi(D_p)=1
\eea
where the product is over the monodromies around the punctures and $a$ and $b$
are the cycles of the torus $T^2$.
We assume translational invariance, that means that the monodromy
around the punctures are constant up to a possible conjugacy transformation
by means of the operator of translations of the quenched theory.
In the twisted case the representation is projective and the constant
monodromies may be conjugated either by the translation operator or
by another twist operator:
\bea
\Psi(a) \Psi(b) \Psi(a)^{-1} \Psi(b)^{-1} \prod_p \Psi(D_p)=\exp(icV^{(2)}) 1
\eea
where $V^{(2)}$ is the area of our space-time torus.
The next version of the derivation of the effective action from the loop 
equations is based on the fact that for $U(N)$ bundles, that we have in any
case to introduce because of the occurrence of the central extension in the
master equation, we can realize explicitly the $R^2$ symmetry as a 
global symmetry, even for the de Rham systems. In fact what really matters to
generate the Cauchy kernel in the loop equations is the relation:
\bea 
\bar{\partial}b-i\mu'=0 
\eea
that follows from Eq.(35) in the gauge $\bar b=0$. Eq.(64) holds under
slightly weaker conditions than $\bar b=0$, that allow us to keep
explicitly a residual $R^2$ global symmetry.
For the Lie algebra $u(N)$ we decompose the connections
$B$ and $b$ into their semi-simple and central parts:
$B=B^s+B^c , b=b^s+b^c$. Our gauge fixing is now:
\bea
\partial \bar b^c=0 \nonumber \\
\bar b^s=0
\eea
This gauge fixing admits the residual global symmetry
$b \rightarrow b+ip_z dz, \bar b \rightarrow \bar b-ip_{\bar z} d\bar z $
maintaining at the same time the validity of Eq.(64).
If this symmetry is unbroken, than the expectation value of the trace of open
$b$ loops vanishes because of the equation:
\bea
 \langle  \Psi(x,y;b^c) \rangle =
 \exp(ip(x-y))
 \langle \Psi(x,y;b^c) \rangle 
\eea
We get:
\bea
0=\int d^2p  D\mu'  D\bar{\mu}' Tr \frac{\delta}{\delta \mu'(w)}
  (\exp(- \Gamma)
  \Psi(x,x;b))
  \eea
where we have inserted the integration over the zero mode $d^2p$.
In the same vein as before we obtain:
\bea
  0=\int d^2p   D\mu'  D\bar{\mu}'\exp(-\Gamma)
  (Tr(\frac{\delta \Gamma}{\delta \mu'(w)} \Psi(x,x;b))+ 
  \nonumber \\ 
  +\int_{C(x,x)} dy_z \bar{\partial}^{-1}(w-y) 
  Tr(\lambda^a \Psi(x,y;b)
  \lambda^a \Psi(y,x;b)) )= \nonumber \\
  =\int d^2p D\mu'  D\bar{\mu}' \exp(-\Gamma)
  (Tr(\frac{\delta \Gamma}{\delta \mu'(w)} \Psi(x,x;b)) +\nonumber \\
  +\int_{C(x,x)} dy_z \bar{\partial}^{-1}(w-y)Tr( \Psi(x,y;b))
    Tr(\Psi(y,x;b))) \nonumber \\
\eea
Then the master equation follows as before. 
Several observations are in order. \\
An important feature of our regularized master equation is that it depends
implicitly on the conformal diffeomorphism that maps the interior of the loop
into the upper half plane. This is a direct consequence of our choice of
regularising the quantum contribution to loop equations in a universal way.
Indeed the local field for which
the loop equations can be reduced to the master equation is the field
$\mu_{t \bar{t}}$ that transforms as a $(1,1)$ form under conformal
diffeomorphism. Instead $\mu_{z \bar{z}}$ is the field that occurs in the
classical action, where $(z,\bar{z})$ are the global coordinates on the torus.
The existence of this conformal mapping plays a key role when we introduce
a lattice divisor, $D$, to get an explicit realization of our Hitchin system,
actually the only one possible if we require a Hausdorff moduli space.
In this case $\mu$ is a collective field given, according to sect.(3)
and appendix B, by:
\bea
\mu^0= \sum_{p} \mu^0_p \delta^{(2)}(x-x_p) \nonumber \\
n= \sum_{p} n_p \delta^{(2)}(x-x_p)\nonumber \\
\bar{n}=\sum_{p} \bar{n}_p \delta^{(2)}(x-x_p)
\eea
Thus $\mu_p$ is a "lattice (or adelic) field of residues".
As a consequence we need a more sophisticated version of the uniformization map.
In this case we ought to map conformally the interior of the loop minus the
internal punctures of the lattice divisor into a punctured disk or into a
fundamental region in the 
upper half-plane. In this case the image of the punctures in the upper
half plane are cusps touching the real axis.
The loop equations in terms of the "lattice field of residues" are thus:
\bea
0=\int \prod_q D\mu_q ' D\bar{\mu_q} ' \exp(-\Gamma)
  (Tr(\frac{\delta \Gamma}{\delta \mu_p}\Psi(x_p,x_p;B)) + \nonumber \\
 +  \int_{C(x_p,x_p)} dy_z \bar{\partial}^{-1}(x_p-y)Tr(\Psi(x_p,y;B))
    Tr(\Psi(y,x_p;B)))
\eea
The cusp singularity at the point $x_p$ introduces one more sign ambiguity
for the central term in the effective action.
In the lattice field version of our loop equations, represented by Eq.(70), we must
take by necessity 
the base-point of the loop at a puncture, otherwise we get a trivial result.
This may seem unnatural from the point of view of the uniformization map,
but somehow surprisingly this is precisely the presentation that occurs
in Penner \cite{Pen} definition of the Teichmuller space of bordered surfaces,
whose moduli involve an arbitrary choice of a point on the boundary,
that is then projected to infinity to become a cusp on the boundary of the 
upper-half plane.
In appendix C we will see how the existence of the conformal diffeomorphism 
involved in the derivation of our master
equation for the field of residues seems to be crucial to get the correct area
law for the Wilson loop in two dimensions by the Hitchin systems. \\
In a sense we may consider the effective action as the image of a
"holographic" map from the boundary, i.e. the loops, to the bulk, i.e. the torus
$T^2$, the space-time of the quenched theory.
From this point of view the HF property of the master algebra is related
to how much entropy is encoded into the boundary. For self-avoiding loops
we have a HF effective action, while for arbitrary self-intersection
of the boundary we may need a non-HF algebra. \\
The preceding derivation of the master equation from the loop equations
literally allows us to employ the master field $B$, that is the
solution of Eq.(70), only to compute the corresponding Wilson loop 
$\lim_N N^{-1}Tr \Psi(x,x;B)$. However the quantity that has 
most physical interest is rather $\lim_N N^{-1}Tr \Psi(x,x;A)$,
i.e. the trace of the unitary monodromy.
This quantity can be computed via a limiting procedure taking advantage of
the existence of the following $\kappa$ action $(\psi \rightarrow \kappa \psi)$
on the Hitchin bundles \cite{Hi1} (see appendix B).  
After some obvious rescaling in the defining equations of our microcanonical
ensemble we get a $\kappa$ dependent effective action appropriate for the
study of the monodromy of the operator 
$B^{ \kappa }=A+i (\kappa \psi+\bar{\kappa} \bar{\psi})= A^{ \kappa }+i {\Psi}^{\kappa} $:
\bea
 Z=\int 
 \delta(F_{ B^{\kappa} }-\mu - \theta^{-1} |\kappa|^2 1) 
 \delta(d^{*}_{A^{\kappa}} {\Psi}^{\kappa}-\nu) \times \nonumber \\
  \times \exp(-S_{TW}) DB  \frac {D (\mu ,\nu)}{D( \mu',\bar{ \mu}')} 
 D \mu'  D \bar{\mu}'=  \nonumber \\
=\int \exp(-\Gamma^{\kappa}) D \mu' D \bar{\mu}'
\eea
After that the monodromy of $A$ is recovered in the limit $\kappa \rightarrow 0$.
This poses the interesting problem as to whether there exist
solutions for $A$ and $\Psi$ that are $\kappa$ independent, as it would be required
by the spirit if not the letter of the master field philosophy, and more 
generally Euclidean (or Lorentz) covariant, in the sense that the Euclidean or
Lorentz group are embedded into the (global) gauge group \cite{W}.
As far as $\kappa$-independence is concerned it is clear that any $\kappa$-independent
solution must coincide with a fixed point of the $\kappa$ action as
$\kappa \rightarrow 0$ on the infinite dimensional Hitchin
bundles involved. The study of the limit $\kappa \rightarrow 0$  would also
furnish 
the Wilson loop for the $A$ monodromy.
We are therefore interested to study the $\kappa$ action on infinite dimensional
Hitchin systems, a case on which we have not much information.
However in finite dimensions such fixed points 
exist and form a Lagrangian subspace of the moduli of Hitchin bundles
that coincides with those bundles that arise as variations of Hodge structures
(see appendix B) \cite{S1,S2,S3}.
For such bundles the $B$ monodromy coincides with the $A$
monodromy \cite{S1,S2,S3,S4} (see appendix B).
As far as Euclidean or Lorentz covariance is concerned, bundles that arise
as variations of Hodge structures are associated to embeddings of $sl(2,R)$
into $sl(N,C)$ via the Schmid nilpotent orbit theorem
\cite{S1,S2,S3}.
It is natural to speculate, taking into account also the peculiar interplay
between the Euclidean and Minkowski version of the master equation, that such
embeddings may be
related to embeddings of the Lorentz group into the gauge group.
In addition we observe that a
necessary condition for Euclidean or Lorentz covariance is given by the
vanishing in the large-$N$ limit of the central extension in the 
microcanonical ensemble for the master field.
Indeed any central element can only transform as a scalar under Euclidean or 
Lorentz transformations, if these are to be embedded into the gauge group,
while this contradicts the nature of the defining equations
of the microcanonical ensemble, that transform as a anti-selfdual tensor in our
four dimensional interpretation. We have already seen that the effective
action develops a central term as result of the quantum contribution.
However this term must be an artifact due to the "chirality" of our change of
variables. Therefore it must be cancelled by the contribution of an
"anomalous" Jacobian of the change of variables in the bulk.
In the next section we will see that $\Gamma$ contains a counter-term
that cancels the quantum contribution provided a suitable finite rescaling of the
conformal factor in the metric is made (indeed the "anomalous" Jacobian depends
on the conformal factor of the metric, once the complex structure is fixed,
while the quantum contribution to the effective action,
being a residue, does not.)

\section{The effective action of large-$N$ $QCD_4$ and of its ${\cal N}=1$ 
extension}

We are now ready to look at the effective action for Hitchin systems in some detail.
It easy to see that the effective action, as it would be computed from the loop
equations without gauge fixing, is the sum of the
classical (quenched or twisted) Yang-Mills action plus the logarithm of a 
number of determinants whose origin is the following one. \\
First of all there are the determinants that arise from the localisation
integral on the microcanonical ensemble, i.e. from integrating the delta 
functionals in Eq.(36,39) over the space of connections:
\bea
 Det'(d_B \wedge)^{-1} Det'(\Re d_B^*)^{-1} Det'(p^* \omega) 
\eea
The first two determinants arise from integrating over the tangent space
to the Hitchin equations. The third one is the pull back, on the moduli that are
not local moduli, of the 
symplectic form on the space of connections. According to the formulas on the 
dimension of the moduli space in appendix B, there are not such non-local moduli
on a torus or on a sphere, once the local moduli fields are fixed, while there
are on Riemann surfaces of higher genus. Therefore for the quenched or twisted
theory on a torus the third factor is absent. 
The $'$ suffix refers to the need of projecting
away from the determinants the zero modes due to gauge invariance, since
gauge-fixing is not implied in the loop equations, though it may be understood
if we like to.
Finally there is the determinant that arises as the Jacobian of the change of
variables from the local moduli fields of the Hitchin system to the 
corresponding holomorphic de Rham local system:
\bea  
Det'\frac{D(\mu ,\nu)}{D(\mu',\bar{\mu}')}
\eea
At formal level and without an explicit gauge fixing, the effective
action, $\Gamma$, reads:
\bea
\Gamma= S_{YM}-logDet'(d_B \wedge)-logDet'(\Re d_B^*)+ 
logDet' \frac{D(\mu ,\nu)}{D(\mu',\bar{\mu}')}
\eea
where $S_{YM}$ is the Yang-Mills action of the reduced quenched or twisted
theory. The effective action can be given a non-formal meaning introducing a metric,
to make sense of the formal determinants, and making a choice of gauge
by the Faddeev-Popov procedure, to avoid the zero modes.
A convenient choice of gauge is \cite{Biq}:
\bea
d_A^* \delta A+i ad_{ \Psi}^* \delta \Psi=0
\eea
that coincides with a background Euclidean or Lorentz covariant gauge
(after analytical continuation to Minkowski space-time).
This gauge fixing condition can be written in terms of $B$ as \cite{Biq}:
\bea
\Re (d_B^*\delta B)=0
\eea
After introducing this gauge-fixing and the corresponding Faddeev-Popov
determinant, the localisation integral over the microcanonical
ensemble gives us:
\bea
Det(-\Delta_1(B))^{-\frac{1}{2}} Det(-\Re \Delta_B)^{-\frac{1}{2}}
Det(-\Im \Delta_B)^{-\frac{1}{2}}Det(-\Re \Delta_B)^{-1}
\eea
where $\Delta_B$ is the scalar Laplacian and $\Delta_1(B)$ is the operator:
\bea
\Delta_1(B)= d_{B}^* d_B \wedge
\eea
acting on one-forms and restricted to $d_B^* \delta B=0$.
The operator $-\Delta_1(B)$ arises from interpreting $\int DB \delta(F_B-\mu)$ as:
\bea
\lim_{\epsilon \rightarrow 0} \int DB \exp(-\frac{1}{\epsilon}
Tr((F_B-\mu)(F_B-\mu)^*))
\eea
with the constraint $d_B^* \delta B=0$. 
The last determinant in Eq.(77) is the Faddev-Popov determinant. The last three factors
in Eq.(77) cancel together. The effective action is thus:
\bea
\Gamma= S_{YM}-\frac{1}{2}logDet(-\Delta_1(B))
 +logDet \frac{D(\mu ,\nu)}{D(\mu',\bar{\mu}')}
\eea
The first determinant, that we obtained from the 
microcanonical localisation, is essentially a one-loop 
contribution in a background field. We should notice that it contributes two
Hermitian polarisations. In fact the connection $B$ is non-Hermitian and thus it is
determined by four real polarisations, but two of them are killed by the conditions:
\bea
\Re (d_B^*\delta B)=0 \nonumber \\
\Im (d_B^*\delta B)=0 
\eea
that come from the gauge-fixing and from the harmonic constraint
\cite{Biq} in the functional
integral.
However this determinant can be more conveniently defined and computed 
by means of a manifestly Euclidean (or Lorentz) invariant metric and of the
representation of the microcanonical ensemble reported at the end of sect.(3).
Though we use now a manifestly Euclidean (or Lorentz) invariant notation, there is
no problem to interpret the following formulas appropriately for the quenched or for the
twisted models.
The localisation integral in Eq.(40) is defined formally as:
\bea
\int DA_{\mu} \delta(F^-_{\mu \nu}-\mu ^-_{\mu \nu})= Det'^{-1}(P^- d_A \wedge)
=Det'^{-\frac{1}{2}}(-\Delta_1)
\eea
where $P^-$ is the projector onto the anti-selfdual part of the field strength.
After inserting the gauge-fixing condition and the corresponding Faddeev-Popov
determinant it can be defined non-formally as:
\bea
Det^{-\frac{1}{2}}(-\Delta_1)= \lim_{\epsilon \rightarrow 0}
 \int Dc DA_{\mu} \exp(-\frac{1}{\epsilon}Tr(c^2)) \times \nonumber \\
\times \exp(-\frac{1}{\epsilon}
Tr((F^-_{\mu \nu}-\mu ^-_{\mu \nu})^2) \delta(D_{\mu} \delta A_{\mu}-c) \Delta_{FP}
\eea
The result for $ Det^{-\frac{1}{2}}(-\Delta_1)$ is thus:
\bea
Det^{-\frac{1}{2}}(-\Delta_1)= Det^{-\frac{1}{2}}(-\Delta_A \delta_{\mu \nu}1+i ad_{F^-_{\mu
\nu}}) \Delta_{FP}
\eea
where $F^-_{\mu\nu}$ is the anti-selfdual part of the field strength, given by:
\bea
F^-_{\mu\nu}=F_{\mu\nu}-F^*_{\mu\nu}
\eea
with:
\bea
F^*_{\mu\nu}=\frac{1}{2} \epsilon_{\mu \nu \alpha \beta} F_{\alpha \beta}
\eea
It is a fundamental problem to study the renormalization group
properties of the master equation for our effective action, with special
regard to the asymptotic freedom \cite{AF,AF1}.
In appendix D we have computed the divergent part of the localisation
determinant $ Det^{-\frac{1}{2}}(-\Delta_1)$ considered as a functional of the
anti-selfdual $F^-_{\mu \nu}$ or self-dual field strength $F^+_{\mu \nu}$:
\bea
-\frac{1}{2}logDet(- \Delta_1)=- \frac{11}{12}\frac{N}{(2 \pi)^2} log(\frac{\Lambda^2}{\mu^2}) \int
d^4x \frac{1}{4}Tr(F^-_{\mu \nu})^2 + \nonumber \\
+ \frac{1}{12}\frac{N}{(2 \pi)^2}  log(\frac{\Lambda^2}{\mu^2}) \int d^4x \frac{1}{4}Tr(F^+_{\mu \nu})^2=
\nonumber \\
=- \frac{12}{12}\frac{N}{(2 \pi)^2} log(\frac{\Lambda^2}{\mu^2}) \int
d^4x \frac{1}{4}Tr(F^-_{\mu \nu})^2 + \nonumber \\
+ \frac{1}{12}\frac{N}{(2 \pi)^2}  log(\frac{\Lambda^2}{\mu^2}) \int d^4x Tr(F_{\mu \nu})^2
\eea
We should notice that the logarithmic divergence associated to
$Tr(F^+_{\mu \nu})^2$ really occurs in the
effective action only if the effective action is regarded as a local functional of
$ F^-_{\mu \nu}$ and $ F^+_{\mu \nu}$. Yet in the master equation
we should rather consider the effective action as a functional
of the moduli fields. Now the components of the  tensor $ F^-_{\mu \nu}$ are certainly
a local functional of
the moduli fields, since the moduli fields are a linear combination of them, while 
$ F^+_{\mu \nu}$ is not, since:
\bea
P^-(d_A \delta A)=\delta \mu ^-
\eea
Because of this non-local character it is not obvious that, once the
non-local expression in terms of the moduli fields is inserted in the definition of the localisation
determinant, it produces a logarithmic divergence.
In any case the master equation reads:
\bea
\frac{\delta \Gamma_{eff}}{\delta \mu(x)}=0
\eea
The left hand side contains, by the definition of the moduli fields $\mu$, a linear
combination of:
\bea
\frac{\delta \Gamma_{eff}}{\delta \mu ^-_{\mu \nu}(x)}
\eea
Using Eq.(88) we get $\int d^4x Tr(F_{\mu \nu})^2= \int d^4x \frac{1}{4} Tr(F^-_{\mu \nu})^2 $ at lowest order
in powers of $ \mu ^-_{\mu \nu} $, assuming Euclidean (or Lorentz) invariance (of the quenched
theory).
It follows that, at first order in the expansion in powers of the moduli fields,
which by the way coincides at this order with the usual perturbative expansion,
the contribution of the self-dual part of the effective action vanishes.
The first coefficient of the beta function is thus exactly reproduced at this
order by the localisation determinant (for the moment we do not include the
contribution of $logDet \frac{D(\mu ,\nu)}{D(\mu',\bar{\mu}')} $ whose computation
we leave for the future):
\bea
\frac{\delta}{\delta \mu(x)}\int d^4x
(\frac{N}{2 \lambda} \frac{1}{4} Tr(F^-_{\mu \nu})^2 
-\frac{11}{12} \frac{N}{(2 \pi)^2} log( \frac {\Lambda^2}{\mu^2}) 
\int d^4x
\frac{1}{4}Tr(F^-_{\mu \nu})^2)=0
\eea
The preceding construction of the effective action of $QCD_4$ extends
easily to the super-symmetric case without scalars. In fact, if only fermions are
present in addition to the gauge fields, they can be integrated out, the result being a Pfaffian
whose logarithm should be added to the effective action. Thus in the ${\cal N}=1$
case, we get for the effective action:
\bea
\Gamma= S_{YM}- \frac{1}{2}logDet(-\Delta_1(B))+
\nonumber \\
+logDet \frac{D(\mu ,\nu)}{D(\mu',\bar{\mu}')}+ logPf(\sigma_{\mu}D_{\mu})
\eea
We must now give an interpretation of the conformal diffeomorphism, due to the 
uniformization map, compatible with the four dimensional interpretation of the
quenching or twisting procedure. Thus the dependence on the conformal
factor
in the first determinant of Eq.(80) will be only through the conformal anomaly, since it
is the determinant of a Laplacian acting on one-forms in four dimensions 
(though the Laplacian looks two dimensional, we have seen that the missing
two dimensional integration is hidden into the definition of the trace
and in the definition of $\Psi$ in the twisted theory or into the dimensionful 
rescaling of the coupling constant and in the quenching prescription for $\Psi$
in the quenched case).
The second determinant in Eq.(80) instead depends explicitly on the conformal factor
of the metric since it is the determinant of an operator acting on scalars.
One of the features that distinguishes the present approach from the
microcanonical localisation, is the contribution from
the Jacobian of the change of variables from Hitchin bundles to the de Rham
systems:
\bea
Det \frac{D(\mu ,\nu)}{D(\mu',\bar{\mu}')}=Det^{-1} \frac{D(\mu' ,\bar{\mu}')}
{D(\mu,\nu)}= Det^{-1} \left( \matrix{ \frac{D \mu'}{D \mu} & \frac{D \mu'}{D \nu} \cr
\frac{D \bar{ \mu}'}{D \nu} & 0 \cr } \right)
\eea
where to get the zero in Eq.(93) we used the relations:
\bea
d_B^*\delta B=i\delta \nu \nonumber \\
d_B \wedge \delta B= \delta \mu
\eea
This determinant is in fact an "anomalous" Jacobian for the action of the two
dimensional complexified gauge group.
This is reminiscent of the fact that the analogous Jacobian that occurs
in the collective field method applied to loop equations arises as an
anomaly in the algebra of loop substitutions \cite{Rajeev}.
More explicitly we can extract from 
$logDet \frac{D(\mu ,\nu)}{D(\mu',\bar{\mu}')}$ a term linear
in $Tr \mu$ as follows:
\bea
Det (\frac{D\mu}{D\mu'}) = Det^{-1} (\frac{D\mu'}{D\mu})= \nonumber \\
 =Det^{-1} (\frac{DG \mu G^{-1}}{D \mu})=Det^{-1}(1+[\mu, G^{-1}\frac{DG}{D\mu}])
\eea
We now use the formula:
\bea
Det(1+A)=\exp Tr log(1+A)
\eea
Since A in our case is a commutator, in finite dimension the
contribution linear in $A$ would vanish. However this is not anymore true
in infinite dimensions, that is our case, since the trace includes integration
over space-time.
We will apply the general theory of regularized determinants to
the case that $A$ is a pseudo-differential operator.
According to the general theory \cite{det, det1}, the definition of regularized 
determinant employs a generalised
noncyclic trace $TR_{reg}$. The definition of $TR_{reg}$ involves a
choice of a positive pseudo-differential operator $Q$ of positive order $q$.
\bea
TR_{reg}(A)=\lim_{\epsilon \rightarrow 0}  Tr(\exp(- \epsilon Q) A)
\eea
or equivalently (the two definitions are related by a Mellin transform):
\bea
TR_{reg}(A)=\lim_{z \rightarrow 0}  Tr(Q^{-z}A)
\eea
$ TR_{reg}$ is a non-cyclic trace, so that, in general, the trace of a commutator
does not vanish.
The renormalised trace is instead:
\bea
TR_{ren}(A)=\lim_{z \rightarrow 0}  Tr(Q^{-z}A-\frac{1}{zq}res(A))
\eea
where $res(A)$ is the Wodzicki residue of a pseudo-differential operator.
It is defined as:
\bea
res(A)=\frac{1}{(2 \pi)^d} \int_M dx \int_{ |p|=1} tra_{-d}(x,p)
\eea
where $a_{-d}(x,p)$ is the homogeneous term of order $-d$ in the momenta, in the
asymptotic expansion of the matrix valued symbol of $A$.
The Wodzicki residue corresponds to extract the coefficient of the
logarithmically divergent term in the symbol expansion of a pseudo-differential
operator. Since usually anomalies arise from finite integrals,
for anomalies the regularized and renormalised traces coincide.
An important formula to compute the regularized trace is:
\bea
TR_{reg}[A,B]=-\frac{1}{q}res([log |Q|,A]B)
\eea
We are now ready to discuss the contribution of the chiral Jacobian
$Det(\frac{D\mu}{D\mu'})$.
We anticipated that in the effective action the central term that comes
from the residue in the quantum contribution to loop equations must
be cancelled by an analogous term from the chiral "anomalous" Jacobian,
since $QCD$ is non-chiral and there is no central term in the defining
classical theory.
This anomalous term in the chiral Jacobian, that is the trace of the term
linear in $\mu$:
\bea
\frac{1}{q}res([log|Q|, G^{-1}\frac{DG}{D\mu}] \mu)
\eea
defines a cocycle, that coincides with the Radul cocycle.
Now the commutator anomaly for point-wise multiplication vanishes in every
dimension but one, where coincides with the (twisted) Radul cocycle.
However our commutator involves an operator, $G^{-1}\frac{DG}{D\mu}$,
whose symbol
is not a local operator, since the functional derivative of the gauge
transformation with respect to $\mu$ involves the inverse Laplacian. 
Thus in our case the commutator anomaly does not vanish in dimension two,
if we take as $Q$ the two dimensional Laplacian, a choice that seems appropriate
in the quenched theory.
Thus we may compute the $u(1)$ part of our chiral anomaly as follows (for the
metric $dz d \bar z$):
\bea
\frac{1}{q}res([log|Q|, G^{-1}\frac{DG}{D\mu}] \mu)= \nonumber \\
=\frac{i}{4}res([log|\partial \bar{\partial}|,(\partial \bar{\partial})^{-1}] \mu)= \nonumber \\
=\frac{i}{2 \pi} \int d^2x  Tr(\mu(x)) 
\eea
From Eq.(103) we see that there exists a finite mismatch of a factor of $\pi$
between the commutator anomaly
and the quantum central term in the effective action. However the commutator
anomaly depends on the choice of the metric, in particular of the conformal
factor of the metric, while the metric and a fortiori its conformal factor
do not affect the residue in the quantum term. Therefore we are able to cancel
the central term in the effective action in the quenched theory by 
a conformal rescaling of the metric of a factor of $\pi$.
It would be interesting to perform the analogous computation in the
twisted theory. In four dimensions a finite conformal rescaling of the metric changes
the effective action only by finite terms due to the four dimensional conformal
anomaly. The same happens in two dimensions through a different mechanism though:
since the classical action depends
on the conformal scale of the metric, a conformal rescaling is equivalent to a
rescaling of
the dimensionful coupling constant (see appendix C).
To be precise the sign of the commutator anomaly in the Euclidean space-time
is fixed, while it is not so for the sign of the quantum contribution,
so that the cancellation between them can actually  occur only
modulo integers, but in fact this is all that is needed, since the 
$\theta$-term in the effective action is defined modulo integers too.

\section{Loop equations and the geometric Langlands correspondence.}

We have seen in the previous section that loop equations admit as solutions
inductive limits of (projective) representations of the fundamental group of
a punctured torus.
In this section we will use the geometric Langlands
correspondence in the sense of Beilinson and Drinfeld \cite{BD,F} (and references
therein) to interpret the master
equation by
operator solutions of the loop equations obtained by the quantisation of the Hitchin
systems.
As we will see the construction that we are referring to is in a certain sense dual
to the one of sect.(4).
Because of quantisation, the HF of the ambient algebra of the master field is
not preserved a priori, while it is kept quantum integrability.
We have seen that the loop equations:
\bea
0=\int  D\mu' D\bar{\mu}' \exp(-\Gamma)
  (Tr(\frac{\delta \Gamma}{\delta \mu(w)}\Psi(x,x;B))+  \nonumber \\
 +\int_{C(x,x)} dy_z \bar{\partial}^{-1}(w-y)Tr(\Psi(x,y;B))
    Tr(\Psi(y,x;B)))
\eea
for self-avoiding loops are implied by the master equation:
\bea
 0=\frac{\delta \Gamma_{eff}}{\delta  \mu(x)}
\eea
However there is a different way of satisfying the loop equations.
We may think that the functional integration in the loop equations
defines a generalised trace $\langle \Omega\mid  \mid\Omega \rangle $ on the field
algebra. Quantisation in presence of
constraints, that in our case are Hitchin self-duality equations on a
punctured torus, is usually performed in two ways.
Either we first quantise the space of connections
\cite{Wea,Wea1} and then we impose
the constraints on the physical
states of the theory or we first solve the constraints and then
we quantise the reduced theory. In our case we must follow the second
alternative because of the localisation over the microcanonical ensemble
in the functional integral.
It is well known that if the moduli are quantised they are interpreted
as differential operators in some representation of the universal enveloping
algebra \cite{Wea,Wea1,Ne1,F1,F2,F3,F4,F5}. Thus a solution of the loop equations
can be found provided we find a
trace $\langle \Omega\mid  \mid\Omega \rangle $ that satisfies:
\bea
 0=\langle \Omega\mid  
  \frac{\delta \Gamma}{\delta \hat{\mu}(w)}\Psi(x,x;\hat B)\mid\Omega \rangle+  \nonumber \\
 + \int_{C(x,x)} dy_z \bar{\partial}^{-1}(w-y)\langle \Omega\mid 
  \Psi(x,y;\hat B)\mid\Omega \rangle 
    \langle \Omega\mid \Psi(y,x;\hat B) \mid\Omega \rangle
\eea
for self-avoiding loops, that certainly happens if:
\bea
 0=\langle \Omega\mid \frac{\delta \Gamma_{eff}}{\delta \hat\mu(x)}
\eea
where we have introduced the $\hat{}$ superscript to indicate that
moduli are now differential operators on a Hilbert space.
This is our new version of the master equation.
In principle we could parameterise the state $\langle \Omega\mid $, that we are looking for, 
by means of the spectrum of the Hamiltonians ${\it \hat{H}}_i$ of
the quantum Hitchin system (see appendix B),
following the technique used in \cite{F1} to solve the Bethe ansatz equations
for the quantum Hitchin Hamiltonians (on a punctured sphere).
The basic idea there is to interpret Hitchin Hamiltonians as elements
of the centre of a Kac-Moody algebra at critical level, in a way that we
recall briefly below (the trace property should be
checked separately, using perhaps the quasi-conformal structure associated
to the critical level \cite{F4,F5}).
Langlands duality is a "far reaching generalisation of Artin-Hasse theory in
number field" \cite{Gin} that loosely speaking associates to every adelic
automorphic representation on a local field a representation of its
Galois group. Beilinson and Drinfeld started a program of understanding
Langlands duality from a geometrical point of view.
From this point of view to the Galois side corresponds a representation of the
fundamental group of a Riemann surface, while to the automorphic side
correspond certain D-modules on the moduli stack of vector bundles
on a Riemann surface. More precisely Beilinson and Drinfeld proved that the
spectrum of the set of commuting operators obtained quantising the Hitchin
Hamiltonians (i.e. the D-modules on the automorphic side)
is parameterised by opers with values in the dual Langlands group \cite{BD}.
To the opers is then associated a representation of the fundamental group via
the monodromy representation (this is the Galois side).
The D-module associated to the quantisation of Hitchin bundles is in fact a
representation of a Kac-Moody algebra of critical level, actually of its
centre \cite{F1,F4,F5}.
It is known that the common eigenstates of the set of Hitchin Hamiltonians
can be obtained by the Bethe ansatz \cite{F1} (see also \cite{Ne1}). \\
It is also clear that the first variation of the effective action computed at
the master field can be considered as an element of the universal enveloping
algebra of the global gauge group, because of translational invariance.
But the universal enveloping algebra of any Lie group can be embedded into
the enveloping algebra of the corresponding Kac-Moody algebra via the
constant matrices.
Therefore we can add to the Hitchin Hamiltonians, i.e. to the centre of the
universal enveloping algebra of the Kac-Moody algebra of critical level, just
one operator in the universal enveloping algebra and still get a commutative
family.
We choose to add precisely the first variation of the effective action
associated to the quantised Hitchin system.
We can then solve the master equation:
\bea
 0=\langle \Omega\mid \frac{\delta \Gamma_{eff}}{\delta \hat \mu(x)}
\eea 
where $ \langle \Omega \mid $ varies over the family of the common eigenstates
of the quantum Hitchin Hamiltonians that are parameterised by the moduli of opers for
the dual Langlands group:
\bea
 \langle \Omega\mid \lambda_i = \langle \Omega\mid {\it \hat{H}}_i
\eea
Thus we come to the conclusion that while in sect.(4) the master
field was parameterised by the moduli of the representations of the
fundamental group with values in the global gauge group, here the vacuum
is parameterised by the moduli of representations in the dual Langlands group.
It is tempting to identify Langlands duality with 't Hooft electric-magnetic
duality since the $U(1)$ subgroups of the gauge group are mapped into each other
by Langlands
duality precisely in the same way Abelian electric-magnetic duality does.
We should mention that there is a natural physical interpretation of this
construction in terms of the conjectured existence of the gauge-field/string
correspondence \cite{Po1,Po2}. The main difference, apart from the one already mentioned
about quantisation, is that if we insist to interpret Hitchin Hamiltonians
as elements of the centre at critical level, there is no associated
conformal vertex algebra, but rather only a quasi-conformal one into which the master field
would live inside \cite{F4,F5}. We should also mention that the construction of sect.(4)
furnishes, after quantisation, two different representations of the master field, related
by a non-unitary similarity transformation. Indeed the master field $b$ in the holomorphic
gauge would be an element of a chiral vertex algebra, while the master field $B$
in the unitary gauge would be an element of a non-chiral one. While this does not affect
the Wilson loop, because of the required trace property of $\langle \Omega\mid  \mid\Omega
\rangle $, some non-holomorphic information (i.e the harmonic metric) is necessary
to compute the effective action $\Gamma_{eff}$.

\section{Conclusions.}

It is known that the von Neumann algebra generated by the master field
of $QCD$ in $d$ dimensions, for $d$ larger or equal to two, or by any equivalent
Eguchi-Kawai reduction to any lower dimension, is a sub-algebra of a Cuntz algebra
with more than one self-adjoint generator, that is not hyper-finite, i.e. it is not
the weak limit of a sequence of finite dimensional matrix algebras. 
Yet functional techniques  have shown that for $QCD_2$ some special 
observables, such as self-avoiding Wilson loops, admit a description in
terms of a master field that generates a hyper-finite (Abelian) von Neumann
algebra. 
We have shown in this paper that for planar self-avoiding Wilson loops,
constructed by the connection $B=(A_z+ \kappa D_u)dz+(A_{\bar z}- \bar \kappa 
\bar{D_u})d\bar z$ with $\kappa \ne 0$ in a two dimensional Eguchi-Kawai model 
equivalent to the large-$N$ limit of $QCD_4$, the large-$N$ loop equations admit
as solution a master monodromy that, assuming existence of the large-$N$ limit, generates
a hyper-finite von Neumann algebra. 
An analogous construction holds for a partially twisted version,
i.e. for the equivalent non-commutative gauge theory.
In particular the master algebra of the partially quenched or twisted
EK model is the limit group algebra of a sequence of (possibly projective)
representations (with values in the complexified gauge group) of 
the fundamental group of an arbitrarily punctured torus. As such it can be 
related to infinite dimensional Hitchin systems and in fact the loop equations
written in terms of the "adelic field of moduli" of the Hitchin systems:
\bea
0=\int \prod_q D\mu_q ' D\bar{\mu_q} ' \exp(-\Gamma)
  (Tr(\frac{\delta \Gamma}{\delta \mu_p}\Psi(x_p,x_p;B)) + \nonumber \\
 +  \int_{C(x_p,x_p)} dy_z \bar{\partial}^{-1}(x_p-y)Tr(\Psi(x_p,y;B))
    Tr(\Psi(y,x_p;B)))
\eea
are shown to be implied, for planar self-avoiding loops, by a 
master equation that determines the critical set of the following effective action:
\bea
\Gamma_{eff}= S_{YM}-\frac{1}{2}logDet(-\Delta_1(B))
 +logDet \frac{D(\mu ,\nu)}{D(\mu',\bar{\mu}')}+ \nonumber \\
 -\epsilon(C)\frac{i}{2} \int Tr \mu d^2x
\eea
Remarkably the first determinant in Eq.(111) can be computed in a manifestly
four dimensional Euclidean (or Lorentz) invariant way and, once inserted into the
master equation, implies the correct value of the one-loop beta function
up to terms of higher order in powers of the moduli fields. \\ 
There is a dual construction, in the sense of the geometric Langlands
correspondence, in which the Hitchin systems are quantised, the moduli
are differential operators and the effective action is an element of the
enveloping algebra of the global gauge group. Thus the loop equations:
\bea
 0=\langle \Omega\mid  
  \frac{\delta \Gamma}{\delta \hat{\mu}(w)}\Psi(x,x;\hat B)\mid\Omega \rangle+  \nonumber \\
 + \int_{C(x,x)} dy_z \bar{\partial}^{-1}(w-y) \langle \Omega\mid
  \Psi(x,y;\hat B)\mid\Omega \rangle
    \langle \Omega\mid \Psi(y,x;\hat B) \mid\Omega \rangle
\eea
are transformed, for self-avoiding loops, into the condition that the master equation annihilates the vacuum:
\bea
 0=\langle \Omega\mid \frac{\delta \Gamma_{eff}}{\delta \hat\mu(x)}
\eea
A possible physical interpretation of this quantised construction is in terms of 
a quasi-conformal (vertex) algebra relative to the conjectured string/gauge-field
duality.

\section{Appendix A}

In this appendix we follow \cite{C}. \\
A von Neumann algebra is a sub-algebra of the algebra of all bounded
operators on a Hilbert space that is closed under the adjoint operation ($^*$)
and under taking limits in the topology of weak convergence,
the weak topology in short.
The dual of a von Neumann algebra is the set of all bounded linear 
functionals over the algebra.
It is usually assumed that the Hilbert space over which the algebra acts
has a countable basis and
that the von Neumann algebra has a separable dual.
Von Neumann double commutant theorem asserts that a von Neumann algebra
coincides with the commutant of its commutant. The commutant is the set of all
bounded operators that commute with all the operators of the algebra.
The centre of an algebra is the set of all the operators of the algebra
that commute with all operators of the algebra. \\
A von Neumann algebra is a factor if its centre is trivial, i.e. a multiple of
the identity. 
Every von Neumann algebra decomposes as a direct integral of factors.
Two von Neumann  algebras are algebraically isomorphic if there exists 
an invertible linear continuous map that preserves the $^*$-algebraic structure. 
An element of the dual of a von Neumann algebra is usually called a state
or a weight on the algebra. It is called positive if it maps positive operators
(i.e operators whose spectrum lies in [0,$\infty$[) into the positive numbers.
A representation of a von Neumann algebra is a linear *-homomorphism
into the algebra of bounded operators on a Hilbert space.
Every positive weight, $w$, on a von Neumann algebra defines a representation
of the algebra via the celebrated GNS representation.
The GNS representation defines the scalar product on the Hilbert space 
of the representation as $\langle A \mid B \rangle=w(A^*B)$ and the 
action of the algebra on the Hilbert space in the obvious way. \\
The basic result in the classification of von Neumann algebras is the
existence of an algebraic invariant of the algebra, the von Neumann
dimension function, that associates to every orthogonal projector, $P$,
of the algebra ($P=P^*,P=P^2$) a non-negative number.
Factors are classified according to the range of values that 
the dimension function may assume: \\
type I: $ 0,1,2,... \infty $ \\
type $II_1$: $[0,1]$ \\
type $II_{\infty}$: $[0,\infty]$ \\
type $III$: $0,\infty$ . \\
A von Neumann algebra is called finite if it is finite the range of its dimension
function.
The theory of von Neumann algebras is a non-commutative far reaching 
generalisation of the integration theory and spectral theorem 
to which it reduces in the commutative case.
Indeed every commutative von Neumann algebra is algebraically isomorphic
to the algebra of all bounded measurable functions $L^{\infty}(X, d\mu)$ for
some measure space $X$ and thus every weight on a commutative von Neumann
algebra is a measure. \\ 
One of the most striking features of the von Neumann algebras is the occurrence
of a continuous dimension.
In fact for type $I$ and type $II$ the dimension function can be identified
with the trace (suitably normalised) of an orthogonal projection.
Therefore type $I$ and $II$ possess semi-finite traces, while type $III$ has no
non-trivial trace.
A trace, $\tau$, is an element of the dual of the algebra that satisfies
$\tau(AB)=\tau(BA)$. If it exist it is unique for a factor up to a
multiplicative constant. \\
After years of efforts since their invention by Murray and von Neumann
a general structure theorem has been achieved in the classification
of the hyper-finite (or amenable or injective) factors with separable dual.
A von Neumann algebra is hyper-finite if it is the closure of a sequence 
of finite dimensional matrix algebras.
Though many von Neumann algebras satisfy this property, there are some
interesting algebras that do not, as the Cuntz algebra with more than one 
self-adjoint generator.
Here it is the complete list of the hyper-finite factors with separable dual.\\
$I_n$: the algebra of matrices of rank $n$.\\
$I_{\infty}$: the algebra of all bounded operators on a Hilbert space
with countable basis.\\
$R$: the unique hyper-finite factor of type $II_1$. $R$ has many equivalent
realizations: as the countable tensor product of $n \times n$ matrices
equipped with the corresponding unique factorized normalised trace;
as a tracial representation of the canonical anti-commutation relations
on an infinite dimensional Euclidean space; as a Krieger factor associated
to an ergodic transformation with an invariant measure. \\
$R_{0,1}= R \times I_{\infty}$: the unique hyper-finite factor of
type $II_{\infty}$. \\
$III_0$: $R_W$, the Krieger factor associated to an ergodic flow with a
quasi-invariant measure. \\
$III_{\lambda}$: $R_{\lambda}$ with $\lambda \ne 0,1$, the Powers factors. \\
$III_1$: $R_{\infty}=R_{\lambda_1}\times R_{\lambda_2}$ with
$\frac{\lambda_1}{\lambda_2}$ not rational, the algebra of canonical
commutation relations in infinite dimensions.\\
In particular the Powers construction of hyper-finite factors of all types
in terms of pairs: (infinite inductive tensor product of matrix algebras, 
factorized state over the tensor product), whose type depends on certain
conditions about the eigenvalue list of the states, turned out to be
fundamental. \\
We now discuss an application of the algebraic approach to the large-$N$
limit that had some relevance for us in the introduction and in sect.(3).
The simplest example of a hyper-finite master algebra is the one generated by
functions of any single self-adjoint operator,
that obviously is a commutative one, equipped with an arbitrary tracial
state.
This may be the algebra of local observables of a one-matrix model
or, less trivially, the sub-algebra generated by the master field for 
a self-avoiding loop of area $A$ in the large $N$-limit of $QCD_2$: 
\bea
\exp(i\sqrt A (a+a^*))
\eea
where $a$ and $a^*$ are elements of the Cuntz algebra.
In this commutative case hyper-finiteness follows from the
Szego limit theorem for self-adjoint operators, that depends essentially
on the fact that Abelian $C^*$-algebras are type I and thus every trace
is uniformly approximately finite dimensional.
In fact the interested reader may find in \cite{Brown} a detailed study
of how the approximation properties of traces carry information about
(tracial) representation theory of von Neumann algebras.

\section{Appendix B}

Hitchin has studied a non-canonical reduction of the four
dimensional self-dual equations:
\bea
F_{\mu \nu}=(^*F)_{\mu \nu}
\eea
to two dimensions, requiring that the four dimensional gauge connection be
constant along two of the four coordinates \cite{Hi1}. \\
The reduction is non-canonical (i.e depending explicitly on a choice
of the coordinate system) since Hitchin identifies the components
of two polarisations of the four-dimensional connection $(A_u, A_{\bar u})$
with the components of a two dimensional one-form 
$(\Psi=\psi+\bar{\psi}=
A_u dz+ A_{\bar u} d\bar z)$, the Higgs field, in order to write the
reduced self-duality equations:
\bea
F_A-i\Psi^2=0\nonumber \\
\bar{\partial}_A \psi=0\nonumber \\
\partial_{A} \bar{\psi}=0
\eea
as a flat condition for the non-Hermitian connection $B=A+i\Psi$:
\bea
F_B=0
\eea
plus a harmonic condition:
\bea
d_A^*\Psi=0
\eea
for the metric $g$ associated to the one form $\Psi$
\cite{Har}:
\bea
\Psi=g^{-1}dg
\eea
Hence the name of harmonic bundles for the Hitchin systems.
The non-canonical nature of the Hitchin reduction matches nicely 
with the use that we have made of it in the partially quenched or twisted 
EK models in sect.(3).
Hitchin has interpreted the reduced self-duality equations as 
the three moment maps of a hyper-Kahler reduction for the
Hamiltonian action of the local gauge group over the cotangent bundle of
unitary connections (here we take the gauge group to be SU(N) or U(N)),
labelled by the pairs $(A, \Psi)$.
Hitchin equations can be generalised to include coadjoint orbits
at points $p$ of a divisor $D$:
\bea
F_A-i\Psi^2= \sum_{p} \mu^0_p \delta^{(2)}(x-x_p)\nonumber \\
\bar{\partial}_A \psi= \sum_{p} n_p \delta^{(2)}(x-x_p)\nonumber \\
\partial_{A} \bar{\psi}=\sum_{p} \bar{n}_p \delta^{(2)}(x-x_p)
\eea
It is known that for fixed eigenvalues of the Hermitian levels, $\mu^0_p$,
the moduli space corresponding to taking the quotient by the action
of the local compact gauge group is hyper-Kahler for $n_p$
nilpotent. Alternatively the moduli space is hyper-Kahler for fixed coadjoint
orbits, i.e. fixed monodromy around the punctures \cite{Biq}. \\
From our point of view, including coadjoint orbits on an arbitrary
divisor is needed to provide a stratification
of the functional integral that be, at the same time, dense in the sense
of distributions and arising from geometry. 
In fact only for a collective field 
of the form:  
\bea
\mu^0= \sum_{p} \mu^0_p \delta^{(2)}(x-x_p) \nonumber \\
n= \sum_{p} n_p \delta^{(2)}(x-x_p)\nonumber \\
\bar{n}=\sum_{p} \bar{n}_p \delta^{(2)}(x-x_p)
\eea
existence of solutions for the microcanonical ensemble of sect.(4)
with a nice (i.e. Hausdorff) moduli space are guaranteed. \\
The corresponding equation for $B$ now reads:
\bea
F_B=\mu^0+n-\bar{n}
\eea
together with the harmonic condition:
\bea
d_A^*\Psi=n+\bar{n}=\nu
\eea
The harmonic bundles defined by the moment map equations,
with coadjoint orbits included, are called tame harmonic bundles
since in this case the eigenvalues of $\psi$ are multi-valued
holomorphic functions with poles at most of order one at the
divisor of the coadjoint orbits.
There are correspondences between tame harmonic bundles and several
other objects of algebraic and topological origin \cite{S1,S2,S3}.
Some of these correspondences play a fundamental role in this paper 
because they allow us to give a precise meaning, on the dense set defined
by Hitchin systems, to the formal change of
variables that occur in the derivation of the master equation 
of sect.(5) from the loop equations.
One way of understanding the existence of such correspondences
comes from the following general result about hyper-Kahler quotients.
As we have seen the hyper-Kahler quotient is obtained imposing the three
moment maps and taking the quotient by the action of the compact gauge
group. In this way we get the moduli space of tame harmonic bundles.
However we can choose one of the three complex structures and 
the associated non-Hermitian moment map and take the quotient under the
action of the complexification of the local gauge group,
getting the same moduli space.
From Eq.(122) it is follows that to a tame harmonic bundle corresponds
a flat connection and thus a local system, i.e. a representation of the
fundamental group of the surface with a divisor.
More precisely a local system is a fibre bundle with locally
constant transition function and sections $v$ given by solutions
of the equations $d_B v=0$ for the flat connection $B$.
The moduli of local systems can be written in terms of the monodromy of the
connection $B$ along the generators of the fundamental group
of the Riemann surface with a divisor, subjected to the consistency
condition:
\bea
\Psi(a) \Psi(b) \Psi(a)^{-1} \Psi(b)^{-1} \prod_p \Psi(D_p)=1
\eea
after taking the quotient by the action of the global gauge group
that acts by conjugation on the monodromies \cite{Ale}. 
In addition to the flat connection it is possible to associate
a flat holomorphic connection with
regular singularities at the punctures:
\bea
\bar{\partial}b'=i\mu'
\eea
This is the holomorphic de Rham local system we refer to in the
paper.
Taking a reduction with respect to another complex structure we get
the Dolbeault moduli space of Higgs bundles defined by:
\bea
\bar{\partial}_A \psi=n
\eea
The Dolbeault moduli do not play any role in sect.(5) but they will
in sect.(6) when we consider a dual interpretation of the master equation
based on the quantisation of Hitchin systems. In particular Hitchin Hamiltonians
are polynomials in $\psi$:
\bea
{\it H}_i=Tr(\psi^i)
\eea 
For completeness we report a formula for the dimension of the
moduli spaces of $SU(N)$ Hitchin bundles on a Riemann surfaces with a divisor
and nilpotent residue \cite{K,K1,K2,K3,K4}.
\bea
dim N= (2g-2)(N^2 -1)+\frac{1}{2}\sum_{p \in D} f_p
\eea
where $f_p=N^2-\sum_p m_p^2$ and $m_p$ the multiplicity of the weights of the
parabolic flag at $p$ \cite{K3}.
We have seen in sect.(5) that for physical purposes, i.e. for calculating
the string tension, it is more convenient 
to compute the unitary monodromy $\Psi(x,x;A)$ on the master field
rather than the $B$ monodromy. We need therefore a way of separating
the field $A$ from $\Psi$ in the loop equations.
The natural way of doing it is to consider the new connection
$B^{\kappa}=A+i(\kappa\psi+\bar{\kappa} \bar{\psi})$ obtained by a suitable rescaling
of the field $\psi$.
The $A$ monodromy can then be recovered in principle in the limit
$\kappa \rightarrow 0$.
We can thus compute the effective action for the microcanonical
ensemble appropriate for $B^{\kappa} $ and try to solve the master 
equation in the limit $\kappa \rightarrow 0$.
However we can do better. As it has been commented in the main text, it is
in the spirit, if not in the letter of the master field idea, that the master
field should be a $\kappa$
independent solution of the master equation up to gauge equivalence.
Thus $A$ and $\Psi$ should be fixed points for the $\kappa$ action up to gauge
equivalence.
In finite dimensions it is known that such fixed points exist and are precisely
those Hitchin bundles that arise as variations of Hodge structures
\cite{S1,S2,S3,S4}.
In addition for such finite dimensional bundles the monodromy of $B^{\kappa} $ is $\kappa$ 
independent and coincides with $\Psi(x,x;A)$ at the fixed point \cite{S4}.

\section{Appendix C}

In this appendix we study the loop equations for self-avoiding loops in
large-$N$ $QCD_2$ and we compare them with non-Abelian localisation. \\
Localisation was discovered by Duistermaat and Heckmann \cite{DH} as a
mean to compute exactly certain integrals by the saddle-point method
despite they may not be Gaussian. \\
It turned out that what was really involved in such integrals was
some sort of cohomology. An infinite dimensional version of DH
localisation due to Bismut \cite{Bis1, Bis2} was employed by Witten \cite{W2}
for computing exactly the partition function and the cohomology ring of $QCD_2$.  \\
For the sphere Witten localisation reduces to the saddle-point
method. Since it depends on equivariant cohomology, Witten localisation
applies to the partition function and to the observables belonging
to the cohomology ring, but it does not apply to all the gauge invariant
observables of $QCD_2$. In particular it does not apply to Wilson loops.
Nevertheless it turns out that for self-avoiding loops on a large sphere
the result derived by large-$N$ functional techniques and the one
obtained were to hold localisation are surprisingly similar though
they do not agree exactly. \\
Indeed there is no reason why they should agree exactly since
there is no reason why localisation should apply to Wilson loops.
Here we use the techniques of sect.(3) and sect.(4) to explain where
the discrepancy may arise. \\
We have seen that our methods for loop equations lead to a result
similar to localisation, in the sense that they transform loop equations for
self-avoiding loops into a saddle-point problem for a certain effective action.
Yet our saddle-point problem is in fact defined over a $U(1)$ extension of
the original theory, since as a consequence of the loop equations a central
extension is present in the master equation.
The functional techniques and two dimensional localisation instead would predict
that there is no central extension at all.
However we observed in the main text that, despite the master equation is a
priori defined over a central extension, we should require that the
central extension vanishes due to a counter-term in the chiral Jacobian
of the change of variables, as it actually happens provided we perform a finite
conformal rescaling of the metric.
In addition we showed that the master equation in our lattice or adelic version
depends on the (regularized) value
of the derivative $\frac{\partial z}{\partial t}(p)$ of the uniformization map
at the base-point $p$ of
the loop, identified as a puncture of the bordered punctured surface internal
to the loop.
For $QCD_2$ our master equation coincides with the saddle-point equation
obtained by localisation, up to the rescaling due to the mentioned derivative
of the uniformization map. \\
Thus it is precisely the existence of the uniformization map the source of the
discrepancy between DH localisation in $d=2$ and the present approach. Indeed we
should remind the reader that the classical action is written in terms of fields
living on the space-time manifold with global coordinates $(z, \bar z)$, while the
functional determinants that arise in the effective action are defined in terms
of fields living on the image of the uniformization map, with coordinates
$(t, \bar t)$.
Unfortunately we do not have a detailed information about the uniformization
map, that by the way is singular at punctures, so that it must be regularized
and suitably interpreted.
What we do here is to find the value of $\frac{\partial z}{\partial t}(p)$
that is implied by the exact
result obtained by functional methods. We find 
$|\frac{\partial z}{\partial t}(p)|^2=N_C $, where
$N_C$ is the number of punctures inside the loop. Localisation
would imply instead $|\frac{\partial z}{\partial t}(p)|^2= 1$, somehow
incorrectly unless the loop contains just one lattice point. \\
Let us present the detailed computation.
For $QCD_2$ it easy to follow the steps that lead to the master 
equation for $QCD_4$. 
The microcanonical ensemble is described by the equation:
\bea
F_A-\mu=0
\eea
that is the moment map at level $\mu$ for the Hamiltonian action
of the local gauge group with respect to the symplectic structure:
\bea
\omega=\int d^2x Tr(\delta A_z \wedge \delta A_{\bar z}) 
\eea
Alternatively Eq.(129) is the zero level condition for the gauge action
on the gauge connection and the coadjoint orbits.
Inserting the resolution of the identity:
\bea
1= \int \delta(F_A-\mu) D\mu
\eea
into the functional integral:
\bea
 0  =  \int DA D \mu
  \exp(-\frac{N}{2 \lambda} \int Tr F_{A}^2 d^2x) \delta(F_A-\mu) 
\eea  
and changing variables in the functional integral from the moduli
field $\mu$ of the unitary connections to the moduli $\mu'$ of the corresponding
holomorphic local systems:
\bea
\bar{\partial}A'_z-i\mu'=0
\eea
we get the effective action $\Gamma$ for the moduli field $\mu'$:
\bea
 0  =  \int D \mu'
  \exp(-\Gamma)
\eea  
where $\Gamma$ is given by:
\bea
\Gamma=\frac{N}{2 \lambda} \int Tr F_{A}^2 d^2x- logDet'(d_A \wedge)+
logDet'(\frac{D\mu}{D\mu'})
\eea
with the first determinant arising from localisation on a given level and
the second determinant from the chiral change of variables (we assume that
we are on a sphere or on a torus so that there are no non-local moduli).
Here the label $'$ for the functional determinants means that the zero modes
coming from gauge invariance must be projected away.
Then, as in sect.(4), the loop equations follow in the form:
\bea
0=\int D\mu'  \exp(-\Gamma)
  (Tr(\frac{\delta \Gamma}{\delta \mu(w)} \Psi(x,x;A)) +\nonumber \\
  + \int_{C(x,x)} dy_z \bar{\partial}^{-1}(w-y)Tr( \Psi(x,y;A)))
\eea
From which we get the master equation:
\bea
0=\frac{\delta \Gamma_M}{\delta \mu(x)}-\frac{i}{2} Ind_C(x) Tr(1)
\eea
$\Gamma_{eff} $ can be computed non-formally by adding the covariant gauge-fixing:
\bea
d_A^* \delta A=0
\eea
with the corresponding Faddeev-Popov determinant and a choice of the metric:
\bea
\Gamma_{eff}=\frac{N}{2 \lambda} \int Tr F_{A}^2 d^2x- 
\frac{1}{2}logDet(-\Delta_1(A))+ \frac{1}{2}logDet(d_A^*d_A)+ \nonumber \\ 
+logDet(\frac{D\mu}{D\mu'})- \epsilon(C)\frac{i}{2}\int Tr \mu d^2x
\eea
However for computational purposes it is more convenient a different choice of
gauge. If our Riemann surface is a sphere, the Hitchin system 
lives over a punctured sphere with punctures forming a uniform lattice (i.e.
rotational invariant). The universal
cover of a punctured sphere is the upper half plane $H$. Our punctured sphere
is then conformally equivalent to a polygon in $H$ with cusps on the
$x$-axis corresponding to the punctures. On $H$ we choose the gauge
$A_y=0$, that leaves, as a residual gauge symmetry, gauge transformations that are
$y$ independent. In the gauge $A_y=0$ the determinant due to localisation
and the Faddeev-Popov determinant are both field independent and
cancel each other. We can use the residual gauge symmetry to fix the gauge
$\mu^{ch}=0$ (the label $ch$ means the non-diagonal part) at the cusps
(i.e the punctures) on the $x$-axis. 
Then the effective action reduces to:
\bea
\Gamma_{eff}= \frac{N}{2\lambda a^2} \sum_i \sum_p h^{(z)i 2}_p +
 \sum_{i \ne j} \sum_p log(h^{(t) i}_p-h^{(t) j}_p)+ \nonumber \\
 + logDet(\frac{Dh^{(t)}}{Dh^{(t)'}})- \epsilon(C)\frac{i}{2}\int Tr h^{(t)} d^2t
\eea
where the $^{(z)}$ and $^{(t)}$ superscript refer to the domain of definition
of the lattice field, $h_p$, and we have set $a=\frac{2 \pi}{\Lambda}$, 
with $a$ the lattice spacing
corresponding to the cutoff $\Lambda$ of the theory, that comes from the product of delta
functions at the same point in the classical action. The last two terms
in Eq.(140) cancel each other (modulo $1$) (up perhaps to terms of higher order
in $ h^{(t)}$) provided the conformal factor
in the metric is rescaled by a factor of $\pi$ (see sect.(5)).
The result coming from localisation would be instead:
\bea
\Gamma= \frac{N}{2\lambda a^2} \sum_p h^{(z) i 2}_p+
 \sum_{i \ne j} \sum_p log(h^{(z)i}_p-h^{(z)j}_p)
\eea
that is similar up to the mentioned discrepancy in the field definitions.
Should we compute the Wilson loop using the master field
obtained by $\Gamma$ rather than $\Gamma_{eff}$ we would get the wrong result
because of the following subtlety. The critical equation for $\Gamma$ is just the
saddle-point equation for the eigenvalues in the $(z, \bar z)$ coordinates.
The master equation for $\Gamma_{eff}$ is the critical equation for
the lattice or adelic field of residues in the coordinates defined by the
uniformization map, i.e. those labelled by $t$, after a conformal rescaling of 
a factor of $\pi$ to cancel
the central term in the effective action.
Let us suppose that we compute the Wilson loop according to the naive cohomological
localisation:
\bea
W(A)=Z^{-1}\int \exp(-\frac{N}{2\lambda a^2} \sum_p h_p ^{i 2}-
 \sum_{i \ne j}log(h_p^i-h_p^j)) \sum_i \exp(i \sum_p h_p^i)
\eea
Assuming translational invariance in the large-$N$ limit we get:
\bea
W(A)=Z^{-1}\int \exp(-\frac{N N_D}{2\lambda a^2} \sum_i h ^{i 2}-
 \sum_{i \ne j} N_D log(h^i-h^j)) \times \nonumber \\
 \times \sum_i \exp(i N_C h^i)
\eea
where $N_D$ is the number of punctures in our space-time manifold and 
$N_C$ the number of punctures internal to our loop $C$. 
By changing variables to $\theta^i= N_C h^i$ we obtain:
\bea
W(A)=Z^{-1}\int \exp(-\frac{N N_D}{2\lambda (N_C a)^2} \sum_i \theta^{i 2}-
 \sum_{i \ne j} N_D log(\theta^i- \theta^j)) \times \nonumber \\
 \times  \sum_i \exp(i \theta^i)
\eea
The corresponding master equation for $\theta$ is:
\bea
\frac{N}{2 \lambda N_C^2 a^2} \theta^{i}=
 \sum_{j;j \ne i} \frac{1}{ \theta^i- \theta^j} 
\eea
This equation should be compared with Eq.(16) that is the correct result obtained
by functional methods:
\bea
\frac{N}{2 \lambda A} \theta^{i}=
 \sum_{j;j \ne i} \frac{1}{ \theta^i- \theta^j} 
\eea
Since $A=N_C a^2 $ we see that Eq.(145) predicts a string tension wrong by a factor
of $N_C$. \\
In our master equation instead the functional derivative
that acts on the classical term is rescaled by the factor of
$|\frac{\partial t}{\partial z}(p)|^2$
while the functional determinants in the effective action are not.
The master equation derived by $\Gamma_{eff}$ now reads:
\bea
\frac{N|\frac{\partial z}{\partial t}(p)|^2}{2\lambda \pi a^2} h^{(t) i }_p=
\sum_{j;j \ne i} \frac{1}{ h^{(t) i }_p- h^{(t) j}_p } 
\eea
where we have included the conformal rescaling of a factor of $\pi$.
We should however write it in terms of the the rotational invariant field
$h^{(z)}_p$:
\bea
\frac{N|\frac{\partial t}{\partial z}(p)|^2}{2 \lambda \pi a^2} h^{(z) i }_p=
\sum_{j;j \ne i} \frac{1}{ h^{(z) i }_p- h^{(z) j}_p } 
\eea
The equation for $\theta^i= N_C h_p^i$ is then:
\bea
\frac{N|\frac{\partial t}{\partial z}(p)|^2}{2 \pi \lambda N_C a^2} \theta^{i}=
 \sum_{j;j \ne i} \frac{N_C}{ \theta^i- \theta^j} 
\eea
that is the correct one provided $|\frac{\partial t}{\partial z}(p)|^2=N_C $.
We can understand naively why $|\frac{\partial t}{\partial z}(p)|^2 $ should be
of order of $N_C$. Let us recall that our manifold has the
topology of a punctured disk, with the boundary of the disk forming the "large
horocycle", while the punctures are regularized by means of small horocycles
around them. Following \cite{Pen,Pen1}, given a metric
on a fatgraph with lengths $l_i$, there is a unique quadratic differential
$q$ such that the residue of $\sqrt q$ at the puncture is $l_i$, the length
of the horocycle around it. We will suppose that this picture extends to
the case that the puncture is on the "large horocycle". We can identify
the differential $\sqrt q$ with $\frac{\partial z}{\partial t}(p)$. After regularization
it is of order of $\frac{l_p}{a}$ that is proportional to the ratio between the
lengths of the "large" and small  horocycles. Its square is proportional 
to the ratio of the corresponding areas. But this is proportional to the number
of punctures $N_C$ inside the "large horocycle".

\section{appendix D}

In this appendix we compute the divergent part of the localisation
determinant. It is convenient to perform the computation in 
an indirect way, by means of a term by term comparison with the usual one-loop
perturbative contribution to the effective action.
For this purpose let us recall the one-loop perturbative result that
is given by:
\bea
\int Dc DA_{\mu} \exp(-\frac{N}{2 \lambda}Tr(c^2)) \exp(-\frac{N}{2 \lambda}
Tr(F_{\mu \nu}^2)) \delta(D_{\mu} \delta A_{\mu}-c) \Delta_{FP}= \nonumber \\
=Det^{-\frac{1}{2}}(-\Delta_A \delta_{\mu \nu}1+2i ad_{F_{\mu \nu}}) Det(-\Delta_A) 
\eea
where we have inserted the gauge-fixing condition and the corresponding
Faddeev- Popov determinant and by an abuse of language we have denoted with $A$
the classical background field in the right hand side of Eq.(150).
The perturbative computation of the one-loop beta function is the result of
two contributions that are independent within logarithmic accuracy \cite{Pol3}.
The orbital contribution gives rise to diamagnetism and to a positive 
term in the beta function:
\bea
log(Det^{-\frac{1}{2}}(-\Delta_A \delta_{\mu \nu}1) Det(-\Delta_A))=
log Det^{-1}(-\Delta_A)= \nonumber \\
= \frac{1}{12} \frac{N}{(2 \pi)^2}log(\frac {
\Lambda^2}{\mu^2}) \int d^4x Tr(F_{\mu \nu})^2
\eea
where it should be noticed the cancellation of two of the four polarisations between
the first factor and the Faddev-Popov determinant.
The spin contribution gives rise to paramagnetism and to a
overwhelming negative term in the beta function \cite{Pol3}:
\bea
\frac{1}{4}Tr(i2 ad_{F_{\mu \nu}} (-\Delta_A )^{-1}
i2 ad_{F_{\mu \nu}} (-\Delta_A )^{-1})= \nonumber \\
=-\frac{12}{12}\frac{N}{(2 \pi)^2} log(\frac {\Lambda^2}{\mu^2})
 \int d^4x Tr(F_{\mu \nu})^2
\eea
Hence the complete result is:
\bea
- \frac{11}{12} \frac{N}{(2 \pi)^2}log(\frac {\Lambda^2}{\mu^2}) 
\int d^4x Tr(F_{\mu \nu})^2
= -\beta_1 log(\frac {\Lambda^2}{\mu^2}) \int d^4x Tr(F_{\mu \nu})^2
\eea
In the case of the localisation determinant we have instead:
\bea
Det^{-\frac{1}{2}}(-\Delta_1)= \lim_{\epsilon \rightarrow 0} \int Dc DA_{\mu} \exp(-\frac{1}{\epsilon}Tr(c^2)) \times \nonumber \\
\times \exp(-\frac{1}{\epsilon}
Tr((F^-_{\mu \nu}-\mu^-_{\mu \nu})^2) \delta(D_{\mu} \delta A_{\mu}-c)
 \Delta_{FP}
\eea
where we have inserted the gauge-fixing condition and the corresponding
Faddeev-Popov determinant
and by an abuse of language we have denoted by $A$ the background field defined by
the condition $ F^-_{\mu \nu}-\mu^-_{\mu \nu}=0$.
The result for $Det^{-\frac{1}{2}}(-\Delta_1)$ is thus:
\bea
Det^{-\frac{1}{2}}(-\Delta_1)=Det^{-\frac{1}{2}}(-\Delta_A \delta_{\mu \nu}1+i ad_{F^-_{\mu
\nu}}) Det(-\Delta_A) 
\eea
The coefficient of the spin contribution
in this case is one half of the one in the perturbative case,
but the spin term involves the anti-selfdual part of the field strength instead of
the field strength itself. The orbital contribution is the same as in the
perturbative case. Thus we have
for the orbital part:
\bea
\frac{1}{12} \frac{N}{(2 \pi)^2}log(\frac {\Lambda^2}{\mu^2}) \int d^4x 
Tr(F_{\mu \nu})^2
=\frac{1}{12} \frac{N}{(2 \pi)^2}log(\frac {\Lambda^2}{\mu^2})
 \int d^4x
\frac{1}{4}Tr(F^-_{\mu \nu})^2+ \nonumber \\
+ \frac{1}{12} \frac{N}{(2 \pi)^2}log(\frac {\Lambda^2}{\mu^2}) 
\int d^4x
\frac{1}{4}Tr(F^+_{\mu \nu})^2
\eea
where we have used the decomposition:
\bea
F_{\mu \nu}=\frac{1}{2} F^-_{\mu \nu}+ \frac{1}{2} F^+_{\mu \nu}
\eea
For the spin part we get:
\bea
\frac{1}{4}Tr(i ad_{F^-_{\mu \nu}} (-\Delta_A )^{-1}
i ad_{F^-_{\mu \nu}} (-\Delta_A)^{-1})= \nonumber \\
=-\frac{12}{12}\frac{N}{(2 \pi)^2} log(\frac {\Lambda^2}{\mu^2}) 
\int d^4x
\frac{1}{4}Tr(F^-_{\mu \nu})^2
\eea
We see that the spin part has the correct factor of 
$2$ once it is written in terms of the anti-selfdual part of the field strength, a fact that we could
have anticipated from start, to arrive at the final answer:
\bea
-\frac{11}{12}\frac{N}{(2 \pi)^2} log(\frac {\Lambda^2}{\mu^2})
 \int d^4x \frac{1}{4}
Tr(F^-_{\mu \nu})^2+
\frac{1}{12}\frac{N}{(2 \pi)^2} log(\frac{\Lambda^2}{\mu^2}) \int d^4x 
\frac{1}{4} Tr(F^+_{\mu \nu})^2
\eea
that is the one reported is sect.(5).

\end{document}